\documentclass[twoside]{article}
\usepackage[a4paper,margin=1.75in]{geometry}

\usepackage{authblk}
\usepackage[utf8]{inputenc}
\usepackage{graphicx}
\usepackage{amssymb,amsfonts,amsmath,amsthm}
\usepackage{commath}
\usepackage{array}
\usepackage{caption}
\usepackage[usenames,dvipsnames,svgnames,table]{xcolor}
\usepackage{dsfont}
\usepackage{framed}
\usepackage{multirow}
\usepackage{algorithm,algorithmic}
\usepackage[hidelinks]{hyperref}

\usepackage{pgfplots}
\usepackage{tikz}
\usepackage{tikzsymbols}
\pgfplotsset{compat=1.14}
\graphicspath{{figs/}}

\newcommand{\p}{\partial}
\newcommand{\ul}{\underline}

\begin{document}
\title{Pricing Financial Derivatives \\ using Radial Basis Function generated \\ Finite Differences with Polyharmonic Splines \\ on Smoothly Varying Node Layouts}
\author{Slobodan Milovanovi\'{c}}
\affil{\small{Department of Information Technology \\ Uppsala University \\ Sweden}}
\date{}
\maketitle
%
%
%
\begin{abstract}
In this paper, we study the benefits of using polyharmonic splines and node layouts with smoothly varying density for developing robust and efficient radial basis function generated finite difference (RBF-FD) methods for pricing of financial derivatives. We present a significantly improved RBF-FD scheme and successfully apply it to two types of multidimensional partial differential equations in finance: a two-asset European call basket option under the Black--Scholes--Merton model, and a European call option under the Heston model. We also show that the performance of the improved method is equally high when it comes to pricing American options. By studying convergence, computational performance, and conditioning of the discrete systems, we show the superiority of the introduced approaches over previously used versions of the RBF-FD method in financial applications.
\end{abstract}
%
%
%

\section{Introduction}
\label{sec1}
Pricing of financial derivatives is one of the core processes in financial markets. When it comes to popularity, options play one of the main roles on that stage of contingent assets, as they offer more freedom to their holders compared to other similar instruments. Unlike futures, options are contracts that grant the right, but not the obligation to buy or sell an underlying asset at a set price on or before a certain date. Option pricing is often a necessary process for making investment decisions, managing risk and calibrating financial models. The prices of these financial instruments are computed by individuals and institutions across the world many times a day using a plethora of methods. Some of those derivatives may be priced analytically, e.g., European options under the assumptions of the famous Black--Scholes--Merton model \cite{black73, merton73}. However, if we consider other types of options, such as American options or basket options, or perhaps we want to use different pricing models (e.g., a local stochastic volatility model) --- in general, we are not able to derive analytical solutions. In those cases, we need to use different numerical methods to approximate the prices of such options. A diverse overview of the numerical schemes used for option pricing can be seen in the BENCHOP project \cite{vonsydow2015benchop, vonsydow2018benchop}. The results of that work illustrate how challenging pricing problems for different methods can be and how important it is to carefully choose and appropriately develop numerical methods in order to build efficient pricing tools.
\par
In this paper, we focus on improvements and adaptations of the Radial Basis Function generated Finite Difference (RBF-FD) methods for pricing multi-asset options and options under multi-factor models. Both of those pricing problems can be formulated as time-dependent multidimensional partial differential equations (PDEs). As a high-order, mesh-free and sparse numerical method from the RBF family, together with the Radial Basis Function Partition of Unity (RBF-PU) method \cite{shcherbakov2016radial, shcherbakov2015radial, safdari2015radial}, RBF-FD shows strong potential when it comes to solving multidimensional PDEs. We develop on top of the previous results of using RBF-FD for financial engineering \cite{milovanovic2018radial, golbabai2016a, kadalbajoo2015an, kumar2015numerical, kadalbajoo2015application, kadalbajoo2013application}, and use important recent advancement of the RBF-FD approximation in other disciplines \cite{bayona2017role, flyer2016on}, to build stable, accurate, and fast solvers for multidimensional PDEs in finance. The main features of the developed solvers address the previous problems of choosing the RBF shape parameter by using polyharmonic splines (PHSs), and instabilities induced by high condition numbers of the differentiation matrices by using node layouts with smoothly varying density.
\par
The remainder of this article is organized as follows. In Section \ref{sec2}, we formulate the RBF-FD method for option pricing problems and motivate using PHSs and node layouts with smoothly varying density. Then, in Section \ref{sec3}, we demonstrate the benefits of the introduced method on a two-asset European call basket option and American put basket option under the Black--Scholes--Merton model, and a European call option under the Heston model. Finally, in Section \ref{sec4}, we draw conclusions and recommend some future research directions.
%
%
%

\section{Radial Basis Function generated Finite Difference Methods}
\label{sec2}
The RBF-FD methods belong to the family of RBF methods. Using the RBF methods for approximating solutions of PDEs dates back to the beginning of the nineties in the previous century \cite{kansa1990multiquadrics2, kansa1990multiquadrics1}. Ever since, these methods have been used in different fields, including computational finance \cite{fasshauer2004using, hon1999radial, pettersson2008improved}. Although the classic RBF methods (also referred to as global RBF methods) possess some desirable properties such as high order convergence and mesh-free domain discretization, they are featured with dense system matrices which in many cases have very large condition numbers. To overcome these weaknesses, several localized RBF approaches with advanced features were introduced, among which RBF-FD \cite{tolstykh2000using, wright2006scattered} and RBF-PU \cite{wendland2002fast} are the most popular and still actively developed.
\par
In order to apply the method, we observe option pricing problems on the truncated computational domain $\Omega\subset \mathds{R}^{d}$ in the following PDE form
\begin{alignat}{2}
\frac{\p}{\p t}u(t,\ul{x}) + \mathcal{L}u(t,\ul{x}) &= 0, &&\ul{x} \in \Omega, \label{eqPDE} \\
{\color{white}\frac{\p}{\p t}u(t,\ul{x}) + }\mathcal{B}u(t,\ul{x}) &= f(t,\ul{x}),\quad &&\ul{x} \in \p \Omega, \label{eqBC} \\
{\color{white}\frac{\p}{\p t}u(t,\ul{x}) + }u(T,\underline{x}) &= g(\underline{x}), &&\ul{x} \in \Omega, \label{eqIC}
\end{alignat}
where $u(t,\ul{x})$ is the option price; $\mathcal{L}$ is the differential operator of the model; $\mathcal{B}$ is the boundary differential operator that with $f(t,\ul{x})$ defines the boundary conditions for the pricing problem; $g(\ul{x})$ is the payoff function; $\ul{x}$ is the spatial variable representing underlying assets or stochastic factors and $t$ is the time variable.
\par
To construct an RBF-FD approximation, we scatter $N$ nodes across the computational domain $\Omega$. For each node $\ul{x}_j$, we define an array of nodes $\mathbf{x}_j$ consisting of $n_j-1$ neighboring nodes and $\ul{x}_j$ itself, and consider it as a stencil of size $n_j$ centered at $\ul{x}_j$. The differential operator $\mathcal{L}$ defined in (\ref{eqPDE})  is approximated in every node  $\ul{x}_j$ as
\begin{equation}
\mathcal{L}u(\ul{x}_j)\approx\sum_{i=1}^{n_j}{w}_{j}^{i}u_j^{i}\equiv \mathbf{w}_ju(\mathbf{x}_j),\quad j=1,\ldots,N,
\label{eqRBFFD}
\end{equation}
where $u_j^{i} \equiv u(\ul{x}_j^i)$ and $\ul{x}_j^i$ is a locally indexed node in $\mathbf{x}_j$, while $\mathbf{w}_j$ is the array of differentiation weights for the stencil centered at $\ul{x}_j$. In the standard RBF-FD methods, the weights ${w}_j^i$ are calculated by enforcing (\ref{eqRBFFD}) to be exact for RBFs centered at each of the nodes in $\mathbf{x}_j$ yielding
\begin{equation}
\label{eqRBFFDmat}
\left[\begin{array}{cccc}
\phi(\|\ul{x}_j^{1}-\ul{x}_j^{1}\|) & \ldots & \phi(\|\ul{x}_j^{1}-\ul{x}_{j}^{n_j}\|)\\
\vdots & \ddots & \vdots\\
\phi(\|\ul{x}_{j}^{n_j}-\ul{x}_j^{1}\|) & \ldots & \phi(\|\ul{x}_{j}^{n_j}-\ul{x}_{j}^{n_j}\|)
\end{array}\right]
\left[\begin{array}{c}
{w}_j^{1}\\
\vdots\\
{w}_{j}^{n_j}
\end{array}\right]=
\left[\begin{array}{c}
\mathcal{L}\phi(\|\ul{x}_{j}-\ul{x}_j^{1}\|)\\
\vdots \\
\mathcal{L}\phi(\|\ul{x}_{j}-\ul{x}_{{j}}^{n_j}\|)
\end{array}\right].
\end{equation}
In theory on RBF interpolation, it is known that (\ref{eqRBFFDmat}) forms a nonsingular system of equations. Therefore, a unique set of weights can be computed for each node. We arrange those weights in a differentiation matrix $L$ in order to build a discrete spatial operator that approximates $\mathcal{L}$. Since $n_j \ll N$, the resulting differentiation matrix is sparse.
\subsection{Smoothly Varying Node Layouts}
\label{secsmooth}
Although the RBF-FD methods are of a mesh-free nature, we observe in practice that the conditioning of the numerical scheme is highly sensitive to the choice of the node layout. Namely, if a discretized computational domain contains non-smooth changes in density of the node layout, it is very likely that the stencils constructed across those areas will have very large condition numbers, and therefore make the entire approximation unstable. Using Cartesian grids with RBF-FD is one way to be safe, but that is far from optimal when it comes to approximation accuracy and it severely limits the adaptive potentials of the method. Some effort has been made to build custom node layouts for pricing basket options in \cite{milovanovic2018radial}, but that approach is hardly generalizable to other problems.
\par
For a successful implementation of RBF-FD methods, we need to be able to quickly generate node layouts with smoothly varying density. According to \cite{fornberg2015fast}, current methods for node scattering can be seen as either \emph{iterative} methods or \emph{advancing front} methods. The most known example of the iterative type are minimal energy distributions, where a repelling force between the nodes is formulated such that it concentrates the nodes in the areas where increased accuracy is needed \cite{flyer2010rotational}. While these methods produce excellent node layouts, in many cases it is computationally inefficient to build node layouts through iterations. Advancing front methods, on the other hand, build node layouts starting from the boundaries until they fill up the domain, which is usually more efficient.
\par
In this work, we focus on the node placing algorithm, introduced in \cite{fornberg2015fast}. The algorithm is of an advancing front type, with an addition of node repelling iterations which further improve node layout quality close to the boundaries. The first ingredients that are necessary for using the algorithm are the boundaries of the computational domain $\Omega$ and the radius function $R(\ul{x})$, which controls density of the node layout on that domain. Next, we define a rectangle that is slightly larger than our computational domain, and we fill it up using an advancing front type basic node placing scheme that can be found in \cite{fornberg2015fast}. Now, the nodes in the rectangle are scattered according to the radius function $R(\ul{x})$. Then, we superpose the node set in the rectangle with the boundary nodes of our computational domain. At this step, we can use the opportunity to also place the nodes of special interest together with the boundary nodes, as these will stay in their positions in the final node layout. This is extremely useful when we are interested to know the solution at a particular point in the computational domain. Following that, we discard all nodes outside the domain boundary and also the ones that are inside the domain, but within a distance of $\frac{1}{2}R(\ul{x})$ from the boundary nodes. Finally, we run $a \in \mathds{N}$ node repel steps on the nodes that are close to the boundary, i.e., $b \in \mathds{N}$ nearest neighbors to each boundary node, in order to smooth out the irregularities in those areas. This is implemented by using a repel force proportional to $r^{-3}$, where $r\in \mathds{R}$ presents the distances between the nodes. The force is set to act between the $b$ neighboring nodes, and it is used in iterations to push the nodes towards the optima of the local potential wells.
\par
One of the great advantages of such smoothly varying node layouts is that we can place the nodes exactly at the points where we want to know the option price without disturbing the smoothness of the layout. This is great for accurate option pricing for the given spot prices of the underlying assets. With traditional node layouts, we need to use interpolation in order to estimate option prices at the desired points, which introduces an additional error and computational load, or we have to disturb the node layout smoothness by placing nodes at the coordinates of interest.
\subsection{Polyharmonic Splines}
Many RBFs (e.g., Gaussian, multiquadric, inverse quadratic) were considered for approximating differential operators in the literature. Although such approximations are featured with great properties, the linear systems of equations that need to be solved in order to obtain the weights $w_j^i$ are often ill-conditioned. Several past works \cite{davydov2011adaptive, fornberg2011stabilization, flyer2012guide, larsson2013stable, fornberg2013stable, flyer2016enhancing} addressed this problem by adding low-order polynomials together with RBFs into the presented interpolation. Moreover, the shape parameter, which is present in most of the RBFs, needs to be chosen carefully in order to have a stable approximation. The problem of choosing the shape parameter for Gaussian-based RBF-FD schemes is thoroughly examined for option pricing problems in \cite{milovanovic2018radial}, but still remains unsolved for general applications.
\par
\begin{figure}[H]
\centering
\input{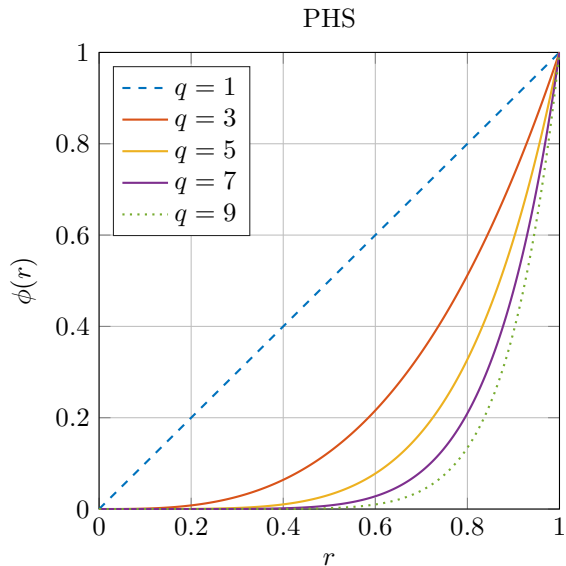}
\caption{Polyharmonic splines of different odd degrees.}
\label{figPHS}
\end{figure}

Nevertheless, recent developments \cite{bayona2017role, flyer2016on}, show that the RBF-FD approximation can be greatly improved by using high order polynomials together with PHSs as RBFs in the interpolation. With that approach, it seems as if the polynomial degree takes the role of controlling the rate of convergence. This allows us to use piecewise smooth PHSs as RBFs without a shape parameter, since the approximation accuracy is no longer controlled by the smoothness of the RBFs. Still, the RBFs do contribute to reduction of approximation errors, and they are necessary in order to have both stable and accurate approximation. We define the PHS function in \eqref{eqPHS} and show some examples in Figure \ref{figPHS},%
\begin{equation}
\label{eqPHS}
\phi(r) =  
	\begin{cases}
		r^q, & q\in\{2k-1\}, \\
		r^q \ln(r), & q\in\{2k\},
	\end{cases}
\end{equation}
where $k \in \mathds{N}$. The results in \cite{flyer2016on} show that there is no significant difference between using odd and even degrees of PHSs in practical applications of RBF-FD. Consequently, we use odd degrees due to their slightly simpler form.
\noindent Taking everything into account, the linear system that we need to solve to obtain the differentiation weights for each node in our problems is
\begin{equation}\label{eq:D2}
\left[\begin{array}{cc}
A & P^T \\
P & 0 \\
\end{array}\right]
\left[\begin{array}{c}
{\mathbf{w}}_j\\
{\mathbf{\gamma}}_j\\
\end{array}\right]=
\left[\begin{array}{c}
\mathcal{L}\phi(\|\ul{x}_{j}-\ul{x}_j^{1}\|)\\
\vdots \\
\mathcal{L}\phi(\|\ul{x}_{j}-\ul{x}_j^{n_j}\|)\\
\mathcal{L}p_1(\ul{x}_j)\\
\vdots\\
\mathcal{L}p_{m_j}(\ul{x}_j)
\end{array}\right],
\end{equation}
where $A$ is the RBF matrix and $\mathbf{w}_j$ is the array of differentiation weights, both shown on the left-hand side of \eqref{eqRBFFDmat}; $P$ is the matrix of size $m_j \times n_j$ that contains all monomials up to degree $p$ (corresponding to $m_j$ monomial terms) that are evaluated in each node $\ul{x}_j^i$ of the stencil $\mathbf{x}_j$ and $\mathbf{0}$ is a zero square matrix of size $m_j \times m_j$; $\mathbf{\gamma}_j$ is the array of dummy weights that should be discarded and $\{p_1, p_2, \ldots, p_{m_j}\}$ is the array of monomial functions indexed by their position relative to the total number of monomial terms $m_j$, such that it contains all the combinations of monomial terms up to degree $p$.
\par
Compared to standard FD discretizations, where differential operators are approximated only on one-dimensional Cartesian grids, meaning that high-dimensional operators need to be discretized separately in each direction, in the RBF-FD approximations dimensionality does not make the problem more difficult. When it comes to the boundary nodes and the nodes that are close to the boundary, the nearest neighbor based stencils automatically form according to the shape of the boundary and require no special treatment for computing the differentiation weights. The only data that is required for approximation of differential operators are Euclidian distances between the nodes. This means that (\ref{eqRBFFDmat}) represents a way to approximate a differential operator in any number of dimensions. Although the FD weights can be directly derived and the RBF-FD weights need to be obtained by solving a small linear system for each node, this task is perfectly parallelizable and that extra cost can be well justified by the desirable features of the method.
\par
After the weights are computed and stored in the differentiation matrix, an approximation of (\ref{eqPDE}) can be presented in the form of the following semi-discrete equation
\begin{align}
\frac{\mathrm{d}}{\mathrm{d} t}\mathbf{u}(t) &= L\mathbf{u}(t), \label{eqdRBFFD}\\
\mathbf{u}(T) &= g(\mathbf{x}),
\end{align}
where $\mathbf{u}(t) \equiv u(t,\mathbf{x})$ is the semi-discrete numerical solution of the pricing equation, while $\mathbf{x}$ is the array of all nodes in the computational domain. To compute the option price $\mathbf{u}(t)$, we need to integrate \eqref{eqdRBFFD} backwards in time.
\subsection{Integration in Time}
For the time discretization we use the second order backward differentiation formula (BDF2) \cite[p.~401]{hairer2000solving}. The BDF2 scheme involves three time levels. To initiate the method, the BDF1 (Euler backward) scheme is often used for the first time step. Thus, two different matrices would need to be factorized. In order to avoid this, we use BDF2 with BDF1 as described in \cite{larsson2008multi}, so that we get a single differentiation matrix.
\par
We split the time interval $[0,T]$ into $M$ non-uniform steps of length $\tau^{l} = t^{M-l}-t^{M-l+1}$, $l = 1,\ldots,M$ and define the BDF2 weights as
\begin{equation}
\beta_0^l = \tau^l\frac{1+\omega_l}{1+2\omega_l},\quad
\beta_1^l = \frac{(1+\omega_l)^2}{1+2\omega_l},\quad
\beta_2^l = \frac{\omega_l^2}{1+2\omega_l},
\end{equation}
where $\omega_l=\tau^l/\tau^{l-1}$, $l=2,\ldots,M$. In \cite{larsson2008multi} it is shown how the time steps can be chosen in such a way that $\beta_0^l\equiv \beta_0$. Therefore, the coefficient matrix is the same in all time steps and only one matrix factorization is needed.
\par
Applying the BDF2 scheme to \eqref{eqdRBFFD} we obtain a fully discretized system of equation
\begin{equation}
(\underbrace{E-\beta_0 L}_{C})
\mathbf{u}^l
= \beta_1^l\mathbf{u}^{l-1} - \beta_2^l\mathbf{u}^{l-2}
\label{impl:system}
\end{equation}
where $E$ is the identity matrix of the appropriate size.

To solve this system, we employ the iterative GMRES method with an incomplete LU factorization as the preconditioner.
%

%

\section{Numerical Experiments}
\label{sec3}
Here, we demonstrate the advantages of the improved RBF-FD method with PHSs on smoothly varying node layouts against the classic RBF-FD setups previously used for option pricing, e.g., in \cite{milovanovic2018radial, vonsydow2015benchop}. We consider three pricing problems. We start with a two-dimensional European call option under the Black--Scholes--Merton model as a simple example. Then, we demonstrate that the method works just as well when used on a more challenging American put basket option under the same model. Finally, as an advanced case, we show the results for a European call option under the local stochastic volatility Heston model.
\par
For all of the considered problems, we scale the operator $\mathcal{L}$ such that we can perform the RBF-FD approximation on a unit domain and then we rescale the result in order to obtain the actual option price. We consider three different node layouts. 
\begin{itemize}
\item The first one is the equidistant Cartesian grid which we refer to as \texttt{cartesian}. 
\item The second layout is the adapted non-uniform node layout used in \cite{milovanovic2018radial} and based on \cite{hout2010adi, haentjens2015adi}, which we call \texttt{adapted}. Conceptually, on a one-dimensional unit domain, we construct this node layout from $N$ equidistant nodes 
$$z_i=\text{arcsinh}\left(-\frac{\hat{K}}{H}\right)+(i-1) \Delta z,$$ 
$i=1,\ldots,N$, where $\hat{K}$ is the scaled strike price position, $H$ is the density parameter, and
$$\Delta z=\frac{1}{N}\left[\text{arcsinh}\left(\frac{1-\hat{K}}{H} \right) -  \text{arcsinh}\left( -\frac{\hat{K}}{H}\right)\right].$$
Then, the final layout would consist of the nodes
$$x_i=\hat{K}+H\cdot \text{sinh}(z_i).$$
In our experiments, we use this one-dimensional concept to construct appropriate node layouts with $H=0.1$. We choose the parameter $H$ empirically, based on several computational experiments. 
\item The third is the newly introduced smoothly varying node layout, constructed by a node placing algorithm from \cite{fornberg2015fast} which we present in Section \ref{secsmooth} and denote here as \texttt{smooth}. We use 
\begin{align}
R(\mathbf{x}) = \frac{1}{\sqrt{N}} \Bigg(&\left(\frac{(\mathbf{x}^{(1)}-X_1)\cos(G)+ (\mathbf{x}^{(2)}-X_2)\sin(G)}{P}\right)^2\nonumber\\
+ &\left(\frac{(\mathbf{x^{(1)}}-X_1)\sin(G) - (\mathbf{x}^{(2)}-X_2)\cos(G)}{Q}\right)^2 + 1\Bigg)\nonumber,
\end{align}
as the radius function, where $\mathbf{x}^{(1)}$ and $\mathbf{x}^{(2)}$ are the components of the vector $\mathbf{x}$, while $X_1$, $X_2$, $P$, $Q$, and $G$ are real parameters intended to control node scattering. As for the local node adjustment, we consider $b=32$ boundary neighbors with $a=4$ repelling iterations. We choose these parameters empirically.
\end{itemize}
By carefully choosing the density controlling parameters, we adapt the former two node layouts to have a higher node density around the discontinuity in the terminal condition, while maintaining desired conditioning and accuracy. 
\par
As RBFs, we use PHSs of degree $q=5$ and augment them with monomials of up to degree $p=4$. This should correspond to an RBF-FD method of fourth order. Nevertheless, since the terminal conditions of the equations that we are solving are not smooth (i.e., the first derivative of the payoff function is discontinuous at the strike price), we can only expect second order convergence. The reason we are not using monomials of degree $p=2$ is that the approximation accuracy and efficiency is still higher with $p=4$, even though both approximations converge with order two. Since the problems are of dimension $D=2$, this gives a polynomial space of size $m=\binom{p+D}{p}=15$, which we use to set the size of the RBF-FD stencils to $n=5m=75$, according to the empirical guidelines from \cite{bayona2017role, flyer2016on}. 
\par
To identify the nearest neighbors for the stencil construction in an efficient way, we employ the $k$-D tree algorithm \cite{bentley1975multidimensional}.
\par
For the plots demonstrating the computational performance, we use \textsc{Matlab} implementations of the presented methods on a laptop equipped with a $2.3$ GHz Intel Core i7 CPU and $16$ GB of RAM. Moreover, the RBF-FD weights computation is performed in parallel using the parallel toolbox command \texttt{parfor} with $4$ workers. We implement the time integration with the \texttt{nofill} setting for the incomplete LU factorization to produce the preconditioner for the GMRES solver. We set the \texttt{tol} parameter in GMRES to $10^{-8}$ for all the experiments. To speed up the convergence, we use the values from the previous time step as the initial value for the next time step. For all experiments, time is discretized into $M=100$ steps. This is just enough to keep the time discretization error smaller than the spatial discretization error in all considered cases.
\subsection{Multi-Asset Options}
A multi-asset option that depends on $D$ underlying risky assets $S_d(t)$, $d=1,\ldots,D$ under the Black--Scholes--Merton model with an assumed risk free bond $B(t)$, follows the dynamics
\begin{equation}\label{eq:multi}
\begin{array}{rcl}
\dif B(t)&=&rB(t)\dif t, \\
\dif S_1(t)&=&\mu_1 S_1(t)\dif t+\sigma_1 S_1(t)\dif W_1(t),\\
\dif S_2(t)&=&\mu_2 S_2(t)\dif t+\sigma_2 S_2(t)\dif W_2(t),\\
\vdots\\
\dif S_D(t)&=&\mu_D S_D(t)\dif t+\sigma_D S_D(t)\dif W_D(t),
\end{array}
\end{equation}
where  $t$ is time, $r$ is the risk free interest rate, $\mu_d$ are the drifts and $\sigma_d$ are the volatilities of $S_d$, and $W_d$ are the Wiener processes. The Wiener processes are correlated such that $\dif W_i(t)\dif W_j(t)=\rho_{i,j}\dif t$. In this multidimensional setting, an option with payoff function $g(S_1(T),\ldots,S_D(T))$, where $T$ is the time of maturity of the option, can be priced under the risk-neutral measure $\mathds{Q}$ as
\begin{equation}
\label{eqMC}
	u(S_1(t),\ldots,S_D(t),t)=e^{-r(T-t)}\mathbb{E}^{{{\mathds{Q}}}}_{t}[g(S_1(T),\ldots,S_D(T))].
\end{equation}
Traditionally, Monte Carlo methods are used to estimate the expected value in \eqref{eqMC} for multi-asset options. In order to apply RBF-FD, we look at the corresponding multidimensional Black--Scholes--Merton equation
\begin{align}
\frac{\partial u}{\partial t}+\mathcal{L}u&=0, \label{eqBS}\\
u(s_1,s_2,\ldots,s_D,T)&=g(s_1,s_2,\ldots,s_D),
\end{align}
where
\begin{equation}
\label{eqBSop}
\mathcal{L}u \equiv r\sum\limits_{i}^{D}s_i\frac{\partial u}{\partial{s_i}}+\frac{1}{2}\sum\limits_{i,j}^{D}\rho_{i,j}\sigma_i\sigma_j s_is_j\frac{\partial^2u}{\partial s_i \partial s_j}-ru,
\end{equation}
and for an arithmetic call option
$$g(s_1,s_2,\ldots,s_D) = \max\left(\frac{1}{D}\sum_{d=1}^D s_d-K,\ 0\right),$$
while for an arithmetic put option
$$g(s_1,s_2,\ldots,s_D) = \max\left(K-\frac{1}{D}\sum_{d=1}^D s_d,\ 0\right),$$
with $K$ as the strike price.
\subsubsection{European Call Basket Option under The Black--Scholes--Merton Model}
\begin{figure}[H]
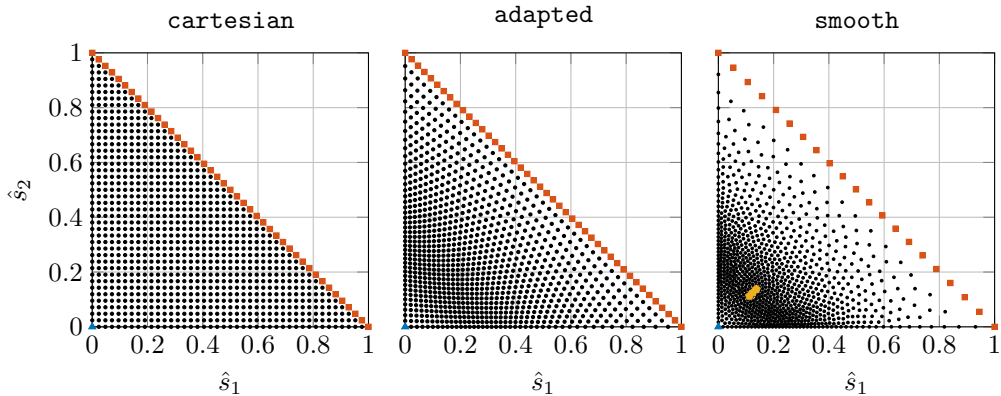

\centering
\input{figs/BSgrid_cart.tikz}
\hspace{0.25cm}
\input{figs/BSgrid_adap.tikz}
\hspace{0.25cm}
\input{figs/BSgrid_smooth.tikz}
\caption{Node layouts on a scaled computational domain used in the numerical experiments. Blue triangles denote the nodes where the close field boundary conditions are applied, red squares denote the nodes where the far field boundary conditions are applied, and yellow pentagons denote the nodes in which the error is measured.}
\label{figBSgrid}
\end{figure}

We perform an experiment with a two-asset European call basket option where $r=0.03$, $\sigma_1=\sigma_2=0.15$, and $\rho=0.5$, with $T=1$ and $K=100$. The error is measured at three points
$$\mathbf{x}_{\texttt{BS}} =
\begin{pmatrix}
90 & 90\\
100 & 100\\
110 & 110
\end{pmatrix},$$ which are close to the strike price $K$. The relative location of these points on a scaled domain can be seen in the right most plot of Figure \ref{figBSgrid} denoted with yellow pentagons. For the \texttt{cartesian} and \texttt{adapted} node layouts, we employ cubic interpolation to approximate the values at the desired points. We then use the maximum error from those three points $\Delta u_{\max}$ to measure convergence.
\par
We set up the computational domain such that the far field boundary at each axis is $s_d^{\max}=8K$ and use the advantages of a mesh-free framework to eliminate the unnecessary computations in the upper right half of the domain by setting the far field boundary diagonally across what would usually be a standard tensor product domain. The node where all underlying assets are equal to $s_d^{\min}=0$, we consider as a close field boundary. There, we set a Dirichlet boundary condition $u(\mathbf{x}_{\texttt{CF}},t)=0$, and at the far field boundary we set $u(\mathbf{x}_{\texttt{FF}},t) = \frac{1}{D}\sum_{i=1}^D s_i-K\exp(-rt)$. An example of the node layouts for this problem where $N\approx1000$ is shown in Figure \ref{figBSgrid}.
\noindent In the plots, the points $\mathbf{x}_{\texttt{CF}}$ are marked by blue triangles, while the $\mathbf{x}_{\texttt{FF}}$ points are marked by red squares. The parameters used to obtain the \texttt{smooth} node layout are $P=0.25$, $Q=0.75$, $G=\frac{\pi}{4}$, while $X_1=X_2=\hat{K}=\frac{1}{8}$.
\par
\begin{figure}[H]
\centering
%
%
\definecolor{mycolor1}{rgb}{0.00000,0.44700,0.74100}%
\definecolor{mycolor2}{rgb}{0.85000,0.32500,0.09800}%
\definecolor{mycolor3}{rgb}{0.92900,0.69400,0.12500}%
\definecolor{mycolor4}{rgb}{0.49400,0.18400,0.55600}%
\definecolor{mycolor5}{rgb}{0.46600,0.67400,0.18800}%
\definecolor{mycolor6}{rgb}{0.30100,0.74500,0.93300}%
\begin{tikzpicture}[trim axis left, trim axis right, baseline]

  \begin{axis}[
  grid=major,
  width=0.46\textwidth,
  height=0.5\textwidth,
  at={(0\textwidth,0\textwidth)},
  scale only axis,
  unbounded coords=jump,
  xmode=log,
  xmin=10^(-2.2),
  xmax=10^(-1.5),
  xlabel={$1/\sqrt{N}$},
  ymode=log,
  ymin=1e-04,
  ymax=1,
  yminorticks=true,
  ytick distance=10^1,
  xminorticks=true,
  ylabel={$\Delta u_{\text{max}}$},
  axis background/.style={fill=white},
  title={Convergence},
  legend pos=north east,
  legend style={legend cell align=left,align=left,draw=white!15!black}
  ]

  \addplot [color=mycolor1,dashed,thick,mark=*,mark options={solid}]
    table[row sep=crcr]{%
    0.0280056016805602	0.206145647906269\\
0.0264350528572715	0.188740764750849\\
0.0250313087160879	0.0305455844783458\\
0.0237691344270764	0.0440491202076103\\
0.0222717701593687	0.0590716726500218\\
0.0209518868547528	0.129808274728419\\
0.0197796943032309	0.0511385279070913\\
0.0187317162316339	0.111572927993496\\
0.0177892016741205	0.0295249497298347\\
0.0167365481751145	0.105371259115391\\
0.0158015154379992	0.0455462316451349\\
0.0149654323603925	0.0821496386864586\\
0.0140719508946058	0.0366143075325525\\
0.0132791469325088	0.0352219101103248\\
0.0125709113788567	0.0224714655097156\\
0.0118345267082788	0.0289513922413978\\
0.011179641181759	0.0121841674644054\\
0.0105934343172574	0.0275811857839097\\
0.00999450453334462	0.0298396100745579\\
0.00939681971469879	0.0157758153698229\\
0.00886658627664886	0.0250910094068253\\
0.0083929957783074	0.00964117408013142\\
0.00792279613759409	0.00690274041388861\\
0.00746289438066075	0.0158440358981631\\
0.00705345615858598	0.0058362078658694\\
};
\addlegendentry{\texttt{cartesian}}

  \addplot [color=mycolor2,dashed,thick,mark=*,mark options={solid}]
    table[row sep=crcr]{%
    0.0274617518190545	0.055800511379517\\
0.0259499648053841	0.0501967431504413\\
0.0245959483971641	0.0238476761216506\\
0.0233762291106092	0.0315581026252536\\
0.0219264504826757	0.0287119439354253\\
0.0206460034753834	0.0188295586978278\\
0.0195068578660218	0.010688966852288\\
0.0184868466661634	0.0161519134581596\\
0.0175682092231577	0.010711338148254\\
0.0165407923394712	0.0190597448605372\\
0.0156269076979498	0.0133106727239873\\
0.0148087219439773	0.00868596533124411\\
0.0139333076031824	0.00982614409359295\\
0.0131556172993976	0.00949415003740928\\
0.012460152291404	0.00813321853102211\\
0.0117363131703255	0.00357404935573879\\
0.0110919563677014	0.0044945402502704\\
0.0105146716312752	0.00625271254385851\\
0.00992436679287576	0.00676420732546124\\
0.00933479382421895	0.00564170298638267\\
0.00881134221062802	0.00309466105906164\\
0.00834347914687139	0.00113904950095289\\
0.00787865765411776	0.0022563411880121\\
0.00742371881148585	0.00182658252606593\\
0.00701845119730874	0.00168287512813414\\
};
\addlegendentry{\texttt{adapted}}
  \addplot [color=mycolor5,dashed,thick,mark=*,mark options={solid}]
    table[row sep=crcr]{%
0.0264906471413009	0.0428274378758404\\
0.025015639663713	0.015035164531314\\
0.0237691344270764	0.00476160050205898\\
0.0226050039454212	0.00468408588252878\\
0.0212285997536077	0.00717295499267045\\
0.0199442341077513	0.00508192185312879\\
0.01898315991505	0.00649951483085154\\
0.0179374000833544	0.00342724899517588\\
0.0171196739349475	0.00347640452786813\\
0.0160872363021947	0.00188019198713008\\
0.0152092354182538	0.00230981460434165\\
0.0144533422340254	0.00199421342329664\\
0.0135768846660426	0.000538281946154129\\
0.0128068376029457	0.00211675572218972\\
0.0121249982949222	0.00163932320348881\\
0.0114489640730411	0.00171331165901734\\
0.010824260542221	0.000585499562291147\\
0.0102614038704477	0.000347423061624497\\
0.00969503978379857	0.00120337969008943\\
0.00912870929175277	0.000963570432130467\\
0.00860790499790018	0.0008628433781771\\
0.00816768882613111	0.000406148034763909\\
0.00773776899975191	0.000830037216140056\\
0.00727123712227554	0.00060958449884474\\
0.00688526292515378	0.00025511827533309\\
    };
    \addlegendentry{\texttt{smooth}}
  \addplot [color=black]
    table[row sep=crcr]{%
    0.0526315789473684	0.621530838558435\\
0.0454545454545455	0.356123331093757\\
0.04	0.68411962566081\\
0.0344827586206897	0.32114971048983\\
0.0303030303030303	0.4716712983092\\
0.0263157894736842	0.252639646767108\\
0.0227272727272727	0.192101125570778\\
0.0196078431372549	0.0898441398589727\\
0.0169491525423729	0.068826271722747\\
0.0147058823529412	0.0817518763796876\\
0.0128205128205128	0.05871301218159\\
0.0111111111111111	0.0479123451714329\\
0.00961538461538462	0.0312883291454309\\
0.00833333333333333	0.0233346593186656\\
0.00719424460431655	0.0108166967389867\\
0.00625	0.0128808967391842\\
0.00543478260869565	0.00960844582551523\\
0.00469483568075117	0.0171164404676691\\
0.00408163265306122	0.0127598891843412\\
0.00353356890459364	0.00206403261669053\\
0.00306748466257669	0.00271903371106852\\
0.00265957446808511	0.00188081756178615\\
0.00230414746543779	0.00131823049766311\\
0.002	0.000839272000262348\\
};
  \legend{};
\end{axis}
\end{tikzpicture}%
\hspace{0.25cm}
%
%
\definecolor{mycolor1}{rgb}{0.00000,0.44700,0.74100}%
\definecolor{mycolor2}{rgb}{0.85000,0.32500,0.09800}%
\definecolor{mycolor3}{rgb}{0.92900,0.69400,0.12500}%
\definecolor{mycolor4}{rgb}{0.49400,0.18400,0.55600}%
\definecolor{mycolor5}{rgb}{0.46600,0.67400,0.18800}%
\definecolor{mycolor6}{rgb}{0.30100,0.74500,0.93300}%
\begin{tikzpicture}[trim axis left, trim axis right, baseline]

  \begin{axis}[
  grid=major,
  width=0.46\textwidth,
  height=0.5\textwidth,
  at={(0\textwidth,0\textwidth)},
  scale only axis,
  unbounded coords=jump,
  xmin=0,
  xmax=30,
  xlabel={time},
  ymode=log,
  ymin=1e-04,
  ymax=1,
  yticklabels={,,}, 
  yminorticks=true,
  ytick distance=10^1,
  xminorticks=true,
  xmajorgrids,
  ymajorgrids,
  axis background/.style={fill=white},
  title={Performance},
  legend pos=north east,
  legend style={legend cell align=left,align=left,draw=white!15!black}
  ]

  \addplot [color=mycolor1,dashed,thick,mark=*,mark options={solid}]
    table[row sep=crcr]{%
1.698083312 0.206145647906269\\
1.724646569 0.188740764750849\\
1.936025464 0.0305455844783458\\
2.100752809 0.0440491202076103\\
2.386483727 0.0590716726500218\\
2.692614552 0.129808274728419\\
2.97423404	 0.0511385279070913\\
3.246785781 0.111572927993496\\
3.583670124 0.0295249497298347\\
4.01425338	 0.105371259115391\\
4.600813218 0.0455462316451349\\
5.123066795 0.0821496386864586\\
5.879724954 0.0366143075325525\\
6.83059396	 0.0352219101103248\\
7.758510705 0.0224714655097156\\
9.038043623 0.0289513922413978\\
9.85679748	 0.0121841674644054\\
14.49073835 0.0275811857839097\\
13.54490130 0.0298396100745579\\
15.32996170 0.0157758153698229\\
17.41498464 0.0250910094068253\\
19.82637399 0.00964117408013142\\
21.83390999 0.00690274041388861\\
25.10149124 0.0158440358981631\\
28.20649025 0.0058362078658694\\
};
  \addlegendentry{\texttt{cartesian}}

  \addplot [color=mycolor2,dashed,thick,mark=*,mark options={solid}]
    table[row sep=crcr]{%
1.637298497	  0.055800511379517\\
1.779654005	  0.0501967431504413\\
1.977625309	  0.0238476761216506\\
2.116564217	  0.0315581026252536\\
2.424778051	  0.0287119439354253\\
2.729700037	  0.0188295586978278\\
3.075075006		0.010688966852288\\
3.369171063		0.0161519134581596\\
3.712687907		0.010711338148254\\
4.240304806		0.0190597448605372\\
4.765269524	 0.0133106727239873\\
5.366471316		0.00868596533124411\\
6.060017341		0.00982614409359295\\
7.118153303		0.00949415003740928\\
7.882492444		0.00813321853102211\\
8.834683471		0.00357404935573879\\
10.125672867	0.0044945402502704\\
14.608040134	0.00625271254385851\\
13.442366422	0.00676420732546124\\
15.464078943	0.00564170298638267\\
17.691689968	0.00309466105906164\\
19.288104951	0.00113904950095289\\
22.24223189		0.0022563411880121\\
25.212779393	0.00182658252606593\\
28.200433636	0.00168287512813414\\
  };
  \addlegendentry{\texttt{adapted}}

  \addplot [color=mycolor5,dashed,thick,mark=*,mark options={solid}]
    table[row sep=crcr]{%
1.936266701	   0.0428274378758404\\
1.974025305	   0.015035164531314\\
2.085315115	   0.00476160050205898\\
2.327457367	   0.00468408588252878\\
2.65599868	   0.00717295499267045\\
2.975735069	 	0.00508192185312879\\
3.191623134	 	0.00649951483085154\\
3.697165251	 	0.00342724899517588\\
4.223008691	 	0.00347640452786813\\
4.509565267	 	0.00188019198713008\\
4.882293879	 	0.00230981460434165\\
5.717093583	 	0.00199421342329664\\
6.194555397	 	0.000538281946154129\\
7.448172418	 	0.00211675572218972\\
8.402410478	 	0.00163932320348881\\
9.211432019	 	0.00171331165901734\\
11.973569763 	0.000585499562291147\\
13.196918576 	0.000347423061624497\\
14.305600677 	0.00120337969008943\\
16.570692576 	0.000963570432130467\\
18.601942577 	0.0008628433781771\\
20.566440029 	0.000406148034763909\\
23.201277741 	0.000830037216140056\\
26.246585618 	0.00060958449884474\\
29.385380653 	0.00025511827533309\\
  };
  \addlegendentry{\texttt{smooth}}
  \addplot [color=black]
    table[row sep=crcr]{%
    0.011210162	0.621530838558435\\
0.015171286	0.356123331093757\\
0.018083279	0.68411962566081\\
0.025154004	0.32114971048983\\
0.032611925	0.4716712983092\\
0.043277852	0.252639646767108\\
0.065305345	0.192101125570778\\
0.08934539	0.0898441398589727\\
0.13174856	0.068826271722747\\
0.19924754	0.0817518763796876\\
0.326996082	0.05871301218159\\
0.515397293	0.0479123451714329\\
0.773311793	0.0312883291454309\\
1.15605595	0.0233346593186656\\
1.94418993	0.0108166967389867\\
3.148218612	0.0128808967391842\\
5.492500073	0.00960844582551523\\
8.582139965	0.0171164404676691\\
13.183536128	0.0127598891843412\\
20.485994551	0.00206403261669053\\
33.149318017	0.00271903371106852\\
55.768370663	0.00188081756178615\\
93.129726603	0.00131823049766311\\
168.119964037	0.000839272000262348\\
};
\end{axis}
\end{tikzpicture}%
\caption{Performance of the RBF-FD method with PHSs for a two-asset European call basket option under the Black--Scholes--Merton model on different node layouts. The left-hand side plot shows the error against the average node layout density and the right-hand side plot shows the error against the computational time measured in seconds. The solid black line represents the performance of the standard FD method of second order.}
\label{figBS}
\end{figure}
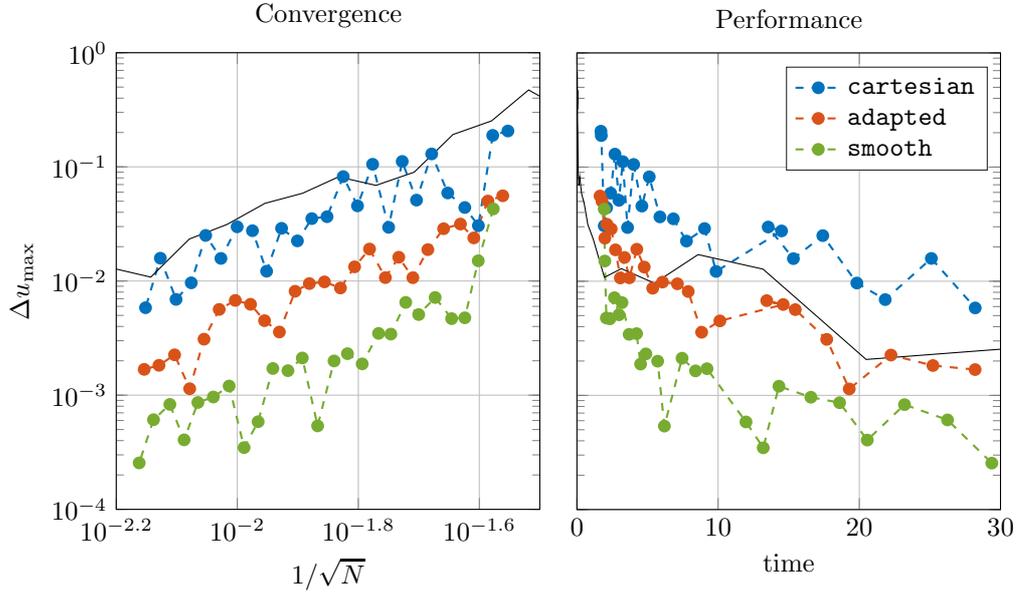

By conducting numerical experiments we have observed second order convergence of the RBF-FD methods as the node layout is refined, which can be seen in Figure \ref{figBS}. From the figure, it is clear that the RBF-FD method on the \texttt{smooth} node layout is significantly more accurate than the other two, for the same amount of computational time. The \texttt{adapted} layout is also performing better than the standard \texttt{cartesian} layout, which is in line with the previous findings presented in \cite{milovanovic2018radial}. In the left-hand side plot of Figure \ref{figBS}, we see that all of the RBF-FD methods are more accurate (relative to the node count) than the standard FD method. This is expected as the \texttt{smooth} layout has higher node density in the area where the error is measured. The right-hand side plot of Figure \ref{figBS} shows that the RBF-FD method on the \texttt{cartesian} node layout can barely compete with the standard FD method, which is reasonable as the computation of the differentiation weights slows it down. Nevertheless, the RBF-FD method evidently outperforms the FD method on the remaining two node layouts.%
\par
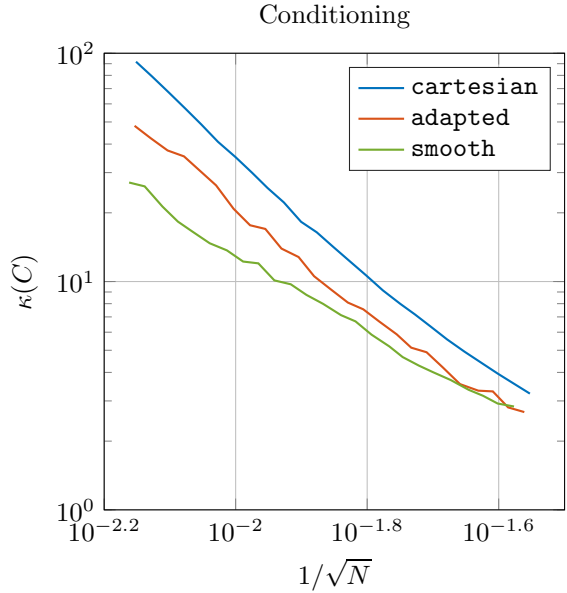
\begin{figure}[H]
\centering
%
%
\definecolor{mycolor1}{rgb}{0.00000,0.44700,0.74100}%
\definecolor{mycolor2}{rgb}{0.85000,0.32500,0.09800}%
\definecolor{mycolor3}{rgb}{0.92900,0.69400,0.12500}%
\definecolor{mycolor4}{rgb}{0.49400,0.18400,0.55600}%
\definecolor{mycolor5}{rgb}{0.46600,0.67400,0.18800}%
\definecolor{mycolor6}{rgb}{0.30100,0.74500,0.93300}%
\begin{tikzpicture}[trim axis left, trim axis right, baseline]

  \begin{axis}[
  grid=major,
  width=0.5\textwidth,
  height=0.5\textwidth,
  at={(0\textwidth,0\textwidth)},
  scale only axis,
  unbounded coords=jump,
  xmode=log,
  xmin=10^(-2.2),
  xmax=10^(-1.5),
  xlabel={$1/\sqrt{N}$},
  ymode=log,
  ymin=10^0,
  ymax=10^2,
  yminorticks=true,
  xtick distance = 10^0.2,
  ylabel={$\kappa(C)$},
  axis background/.style={fill=white},
  title={Conditioning},
  legend pos=north east,
  legend style={legend cell align=left,align=left,draw=white!15!black}
]
\addplot [color=mycolor1,thick]
  table[row sep=crcr]{%
0.0280056016805602	3.22959550067135\\
0.0264350528572715	3.58070256436426\\
0.0250313087160879	3.95008879944081\\
0.0237691344270764	4.35715682632609\\
0.0222717701593687	4.9309820883814\\
0.0209518868547528	5.5820111609325\\
0.0197796943032309	6.35821537858085\\
0.0187317162316339	7.17496711279785\\
0.0177892016741205	7.98889781304042\\
0.0167365481751145	9.14882782202149\\
0.0158015154379992	10.6156465995235\\
0.0149654323603925	12.1520822337279\\
0.0140719508946058	14.1743612226013\\
0.0132791469325088	16.432019506276\\
0.0125709113788567	18.2673327179379\\
0.0118345267082788	22.2106331905467\\
0.011179641181759	25.684280520114\\
0.0105934343172574	29.9257915247501\\
0.00999450453334462	35.0650981803253\\
0.00939681971469879	40.9338540973059\\
0.00886658627664886	48.7251877977137\\
0.0083929957783074	57.0175508618455\\
0.00792279613759409	67.0620428343098\\
0.00746289438066075	79.0727022550011\\
0.00705345615858598	91.8214508020705\\
};
\addlegendentry{\texttt{cartesian}}

\addplot [color=mycolor2,thick]
  table[row sep=crcr]{%
0.0274617518190545	2.68104914272417\\
0.0259499648053841	2.8102343395508\\
0.0245959483971641	3.30246139997641\\
0.0233762291106092	3.32676231822933\\
0.0219264504826757	3.55467135548407\\
0.0206460034753834	4.20601151866921\\
0.0195068578660218	4.90185643789292\\
0.0184868466661634	5.13251296421724\\
0.0175682092231577	5.8736383627833\\
0.0165407923394712	6.67074882568284\\
0.0156269076979498	7.55966831911625\\
0.0148087219439773	8.08917572743865\\
0.0139333076031824	9.26369253258599\\
0.0131556172993976	10.5522995801605\\
0.012460152291404	12.8047757616335\\
0.0117363131703255	13.9436980049024\\
0.0110919563677014	17.0039955115346\\
0.0105146716312752	17.6414286208347\\
0.00992436679287576	20.8339411738284\\
0.00933479382421895	26.3825312976943\\
0.00881134221062802	30.6588530798516\\
0.00834347914687139	35.3421900834912\\
0.00787865765411776	37.5156141202895\\
0.00742371881148585	42.4648915734771\\
0.00701845119730874	48.0931441770536\\
};
\addlegendentry{\texttt{adapted}}

\addplot [color=mycolor5,thick]
  table[row sep=crcr]{%
0.0264906471413009	2.83599374731246\\
0.025015639663713	2.92212772093073\\
0.0237691344270764	3.15963663568794\\
0.0226050039454212	3.35532085321617\\
0.0212285997536077	3.70321224949756\\
0.0199442341077513	4.0106220143046\\
0.01898315991505	4.28471051484707\\
0.0179374000833544	4.67006277239386\\
0.0171196739349475	5.19936526696152\\
0.0160872363021947	5.85249599739638\\
0.0152092354182538	6.69225976993601\\
0.0144533422340254	7.12880912434184\\
0.0135768846660426	7.98645729808365\\
0.0128068376029457	8.74268657600063\\
0.0121249982949222	9.72712583076303\\
0.0114489640730411	10.1161178354606\\
0.010824260542221	12.0137831898859\\
0.0102614038704477	12.234360871873\\
0.00969503978379857	13.6946208414836\\
0.00912870929175277	14.7144084307737\\
0.00860790499790018	16.4952942548323\\
0.00816768882613111	18.3013312243309\\
0.00773776899975191	21.3023989508779\\
0.00727123712227554	26.1159840891661\\
0.00688526292515378	27.1466789928244\\
};
\addlegendentry{\texttt{smooth}}

\end{axis}
\end{tikzpicture}%
\caption{The condition number of the differentiation matrix $C$ as a function of the average node density for different node layouts when pricing a two-asset European Call basket option.}
\label{figBScond}
\end{figure}

From the past experience of working with the RBF-FD methods, we also observe the condition number of the differentiation matrix in order to be able to anticipate potential numerical instabilities. The condition numbers for the three setups are presented in Figure \ref{figBScond}. This figure also shows the benefits of using the \texttt{smooth} layouts versus the others, although the \texttt{adapted} layout is also showing a significantly lower condition number growth compared to \texttt{cartesian} as the node density increases.
\subsubsection{American Put Basket Option under The Black--Scholes--Merton Model}
When it comes to American options, these financial derivatives can be exercised at any $t \leq T$, as opposed to the European options that can only be exercised at $t=T$. Instead of using a PDE as a model, for American options we need to formulate the pricing task as a linear complementarity problem (LCP)
\begin{align}
\frac{\partial u}{\partial t}+\mathcal{L}u&\geq 0,\nonumber \\
u(s_1,s_2,\ldots, s_D,t)&\geq g(s_1,s_2,\ldots, s_D),\label{eqlcp}\\
\left( \frac{\partial u}{\partial t}+\mathcal{L}u\right)& \left(u(s_1,s_2,\ldots, s_D,t)-g(s_1,s_2,\ldots, s_D)\right)=0\nonumber,
\end{align}
with the initial data $g(s_1,s_2,\ldots, s_D)$. In order to solve \eqref{eqlcp}, we use the operator splitting method \cite{ikonen2004operator, ikonen2009operator, salmi2014imex}, combined with the RBF-FD and BDF2 methods, in the same way as it was used in \cite{milovanovic2018radial}.
\par
We conduct this experiment using the same parameters as for the European call basket option. The only difference are the boundary conditions, because now we deal with a put option. The boundary condition at the far-field boundary nodes is set to be $u(\mathbf{x}_{\texttt{FF}},t)=0$. The option price at the close-field boundary is kept at the value of the payoff function at that point, since that node is behind the free boundary characteristic for American options. 
\begin{figure}[H]
\centering
%
%
\definecolor{mycolor1}{rgb}{0.00000,0.44700,0.74100}%
\definecolor{mycolor2}{rgb}{0.85000,0.32500,0.09800}%
\definecolor{mycolor3}{rgb}{0.92900,0.69400,0.12500}%
\definecolor{mycolor4}{rgb}{0.49400,0.18400,0.55600}%
\definecolor{mycolor5}{rgb}{0.46600,0.67400,0.18800}%
\definecolor{mycolor6}{rgb}{0.30100,0.74500,0.93300}%
\begin{tikzpicture}[trim axis left, trim axis right, baseline]

  \begin{axis}[
  grid=major,
  width=0.46\textwidth,
  height=0.5\textwidth,
  at={(0\textwidth,0\textwidth)},
  scale only axis,
  unbounded coords=jump,
  xmode=log,
  xmin=10^(-2.2),
  xmax=10^(-1.5),
  xlabel={$1/\sqrt{N}$},
  ymode=log,
  ymin=1e-04,
  ymax=1,
  yminorticks=true,
  ytick distance=10^1,
  xminorticks=true,
  ylabel={$\Delta u_{\text{max}}$},
  axis background/.style={fill=white},
  title={Convergence},
  legend pos=north east,
  legend style={legend cell align=left,align=left,draw=white!15!black}
  ]
  \addplot [color=mycolor1,dashed,thick,mark=*,mark options={solid}]
    table[row sep=crcr]{%
    0.0280056016805602	0.368235976874585\\
    0.0264350528572715	0.180242959580026\\
    0.0250313087160879	0.0494977567400241\\
    0.0237691344270764	0.102038874060606\\
    0.0222717701593687	0.0309630829080954\\
    0.0209518868547528	0.103918633881213\\
    0.0197796943032309	0.0381678599000317\\
    0.0187317162316339	0.105948206840419\\
    0.0177892016741205	0.0316580268392173\\
    0.0167365481751145	0.19385825270264\\
    0.0158015154379992	0.0397507910360453\\
    0.0149654323603925	0.170772293148891\\
    0.0140719508946058	0.039088604650614\\
    0.0132791469325088	0.0531027927032\\
    0.0125709113788567	0.0718000486264092\\
    0.0118345267082788	0.0287388286474162\\
    0.011179641181759	0.0251805288708891\\
    0.0105934343172574	0.0187218845034947\\
    0.00999450453334462	0.025506020668578\\
    0.00939681971469879	0.018633110738413\\
    0.00886658627664886	0.0411177591061254\\
    0.0083929957783074	0.00598307898769023\\
    0.00792279613759409	0.0114310100251824\\
    0.00746289438066075	0.0111941042012429\\
    0.00705345615858598	0.0280177985694081\\
};
\addlegendentry{\texttt{cartesian}}

  \addplot [color=mycolor2,dashed,thick,mark=*,mark options={solid}]
    table[row sep=crcr]{%
    0.0274617518190545	0.122786798048862\\
    0.0259499648053841	0.0890019840490979\\
    0.0245959483971641	0.00907582117293959\\
    0.0233762291106092	0.123503588221544\\
    0.0219264504826757	0.00716850214012688\\
    0.0206460034753834	0.0391195656913101\\
    0.0195068578660218	0.0220305114074932\\
    0.0184868466661634	0.0185857799236295\\
    0.0175682092231577	0.0579839997451312\\
    0.0165407923394712	0.0190442220654354\\
    0.0156269076979498	0.0136693543207773\\
    0.0148087219439773	0.0401656827807617\\
    0.0139333076031824	0.0178045282629089\\
    0.0131556172993976	0.0141765659115887\\
    0.012460152291404	0.0157853457157895\\
    0.0117363131703255	0.0116389077822827\\
    0.0110919563677014	0.00961281491873578\\
    0.0105146716312752	0.00912463655091322\\
    0.00992436679287576	0.00836817493198261\\
    0.00933479382421895	0.00505111291925209\\
    0.00881134221062802	0.00298973157179816\\
    0.00834347914687139	0.00131181883318554\\
    0.00787865765411776	0.00211004205682208\\
    0.00742371881148585	0.00185255350529623\\
    0.00701845119730874	0.00192990104972601\\
};
\addlegendentry{\texttt{adapted}}
  \addplot [color=mycolor5,dashed,thick,mark=*,mark options={solid}]
    table[row sep=crcr]{%
    0.0264906471413009	0.0278550173199132\\
    0.025015639663713	0.0178938759043641\\
    0.0237691344270764	0.0121334515814535\\
    0.0226050039454212	0.00319129701837362\\
    0.0212285997536077	0.0129635648779889\\
    0.0199442341077513	0.00725642037845819\\
    0.01898315991505	0.00503132589326505\\
    0.0179374000833544	0.00668830617410876\\
    0.0171196739349475	0.00228014702004309\\
    0.0160872363021947	0.00167869642766716\\
    0.0152092354182538	0.0011992489488486\\
    0.0144533422340254	0.00159022754662863\\
    0.0135768846660426	0.00667899990055787\\
    0.0128068376029457	0.00900564264773251\\
    0.0121249982949222	0.0015483215912937\\
    0.0114489640730411	0.00422749874484118\\
    0.010824260542221	0.00383816726280273\\
    0.0102614038704477	0.00195097824642332\\
    0.00969503978379857	0.00340533123181608\\
    0.00912870929175277	0.000595405010821226\\
    0.00860790499790018	0.00083476036727259\\
    0.00816768882613111	0.00110513111745847\\
    0.00773776899975191	0.000704771889217758\\
    0.00727123712227554	0.000408655004219338\\
    0.00688526292515378	0.000500934316051971\\
    };
    \addlegendentry{\texttt{smooth}}

    \addplot [color=black]
      table[row sep=crcr]{%
      0.0526315789473684	0.631672110526619\\
0.0454545454545455	0.540983624152989\\
0.04	1.83097456260111\\
0.0344827586206897	1.42046425888732\\
0.0303030303030303	1.11249692739571\\
0.0263157894736842	0.366877085718592\\
0.0227272727272727	0.360299283349913\\
0.0196078431372549	0.187429903389282\\
0.0169491525423729	0.217476142897075\\
0.0147058823529412	0.195208464502675\\
0.0128205128205128	0.0908479784556797\\
0.0111111111111111	0.126104701432212\\
0.00961538461538462	0.0965788987389349\\
0.00833333333333333	0.042790758012452\\
0.00719424460431655	0.0321681002346068\\
0.00625	0.0347644967256162\\
0.00543478260869565	0.0189407929242655\\
0.00469483568075117	0.0209047176171655\\
0.00408163265306122	0.0155472947377611\\
0.00353356890459364	0.00770146510384784\\
0.00306748466257669	0.00665267595175045\\
0.00265957446808511	0.00438959115551718\\
0.00230414746543779	0.00250498018612522\\
0.002	0.00160657495996475\\
      };
  \legend{};
\end{axis}
\end{tikzpicture}%
\hspace{0.25cm}
%
%
\definecolor{mycolor1}{rgb}{0.00000,0.44700,0.74100}%
\definecolor{mycolor2}{rgb}{0.85000,0.32500,0.09800}%
\definecolor{mycolor3}{rgb}{0.92900,0.69400,0.12500}%
\definecolor{mycolor4}{rgb}{0.49400,0.18400,0.55600}%
\definecolor{mycolor5}{rgb}{0.46600,0.67400,0.18800}%
\definecolor{mycolor6}{rgb}{0.30100,0.74500,0.93300}%
\begin{tikzpicture}[trim axis left, trim axis right, baseline]

  \begin{axis}[
  grid=major,
  width=0.46\textwidth,
  height=0.5\textwidth,
  at={(0\textwidth,0\textwidth)},
  scale only axis,
  unbounded coords=jump,
  xmin=0,
  xmax=30,
  xlabel={time},
  ymode=log,
  ymin=1e-04,
  ymax=1,
  yticklabels={,,}, 
  yminorticks=true,
  ytick distance=10^1,
  xminorticks=true,
  xmajorgrids,
  ymajorgrids,
  axis background/.style={fill=white},
  title={Performance},
  legend pos=north east,
  legend style={legend cell align=left,align=left,draw=white!15!black}
  ]
  \addplot [color=mycolor1,dashed,thick,mark=*,mark options={solid}]
    table[row sep=crcr]{%
    1.266632092	0.368235976874585\\
    1.693104587	0.180242959580026\\
    1.979143504	0.0494977567400241\\
    2.151952013	0.102038874060606\\
    2.391378973	0.0309630829080954\\
    2.643520992	0.103918633881213\\
    3.020969694	0.0381678599000317\\
    3.267646546	0.105948206840419\\
    3.676803846	0.0316580268392173\\
    4.139253406	0.19385825270264\\
    4.721022914	0.0397507910360453\\
    5.300093837	0.170772293148891\\
    6.133775727	0.039088604650614\\
    6.975241242	0.0531027927032\\
    7.750557845	0.0718000486264092\\
    8.698607314	0.0287388286474162\\
    10.060603529	0.0251805288708891\\
    11.497422104	0.0187218845034947\\
    13.308454576	0.025506020668578\\
    15.274714644	0.018633110738413\\
    16.902411434	0.0411177591061254\\
    19.568326814	0.00598307898769023\\
    21.815610615	0.0114310100251824\\
    24.86870614	0.0111941042012429\\
    28.033236335	0.0280177985694081\\
  };
  \addlegendentry{\texttt{cartesian}}

  \addplot [color=mycolor2,dashed,thick,mark=*,mark options={solid}]
    table[row sep=crcr]{%
    1.518264709	0.122786798048862\\
    1.814046741	0.0890019840490979\\
    2.004759125	0.00907582117293959\\
    2.168029261	0.123503588221544\\
    2.475958932	0.00716850214012688\\
    2.741068669	0.0391195656913101\\
    3.09606853	0.0220305114074932\\
    3.356777134	0.0185857799236295\\
    3.803247331	0.0579839997451312\\
    4.177844039	0.0190442220654354\\
    4.800236524	0.0136693543207773\\
    5.335089708	0.0401656827807617\\
    6.048323618	0.0178045282629089\\
    6.99292085	0.0141765659115887\\
    7.780686548	0.0157853457157895\\
    8.975884405	0.0116389077822827\\
    10.149642629	0.00961281491873578\\
    11.736299675	0.00912463655091322\\
    13.239164268	0.00836817493198261\\
    15.526647021	0.00505111291925209\\
    17.6349634	0.00298973157179816\\
    19.397080193	0.00131181883318554\\
    22.293481509	0.00211004205682208\\
    25.16257304	0.00185255350529623\\
    27.788962072	0.00192990104972601\\
  };
  \addlegendentry{\texttt{adapted}}

  \addplot [color=mycolor5,dashed,thick,mark=*,mark options={solid}]
    table[row sep=crcr]{%
    1.582034489	0.0278550173199132\\
    1.953913143	0.0178938759043641\\
    2.35925907	0.0121334515814535\\
    2.411497698	0.00319129701837362\\
    2.686311259	0.0129635648779889\\
    2.968548585	0.00725642037845819\\
    3.324908278	0.00503132589326505\\
    3.740409163	0.00668830617410876\\
    4.074877007	0.00228014702004309\\
    4.570754442	0.00167869642766716\\
    5.124059612	0.0011992489488486\\
    5.735901262	0.00159022754662863\\
    6.430426764	0.00667899990055787\\
    7.362663247	0.00900564264773251\\
    8.335174883	0.0015483215912937\\
    9.640431203	0.00422749874484118\\
    11.118793033	0.00383816726280273\\
    12.653374268	0.00195097824642332\\
    14.318903827	0.00340533123181608\\
    16.55403856	0.000595405010821226\\
    18.822081091	0.00083476036727259\\
    20.391740522	0.00110513111745847\\
    23.477533895	0.000704771889217758\\
    26.760489015	0.000408655004219338\\
    28.482961009	0.000500934316051971\\
  };
  \addlegendentry{\texttt{smooth}}
  \addplot [color=black]
    table[row sep=crcr]{%
    0.026579247	0.631672110526619\\
0.012510452	0.540983624152989\\
0.013996884	1.83097456260111\\
0.020530223	1.42046425888732\\
0.041039322	1.11249692739571\\
0.038814532	0.366877085718592\\
0.059881837	0.360299283349913\\
0.091762291	0.187429903389282\\
0.145278834	0.217476142897075\\
0.233736007	0.195208464502675\\
0.359191046	0.0908479784556797\\
0.576829172	0.126104701432212\\
0.981836382	0.0965788987389349\\
1.614060888	0.042790758012452\\
2.796724445	0.0321681002346068\\
5.031241597	0.0347644967256162\\
8.366985585	0.0189407929242655\\
14.0369007	0.0209047176171655\\
23.793249886	0.0155472947377611\\
40.948639945	0.00770146510384784\\
71.541796825	0.00665267595175045\\
124.559925346	0.00438959115551718\\
215.687845368	0.00250498018612522\\
380.920992257	0.00160657495996475\\
    };
\end{axis}
\end{tikzpicture}%
\caption{Performance of the RBF-FD method with PHSs for a two-asset American put basket option under the Black--Scholes--Merton model on different node layouts. The left-hand side plot shows the error against the average node layout density and the right-hand side plot shows the error against the computational time measured in seconds. The solid black line represents the performance of the standard FD method of second order.}
\label{figBSamPut}
\end{figure}
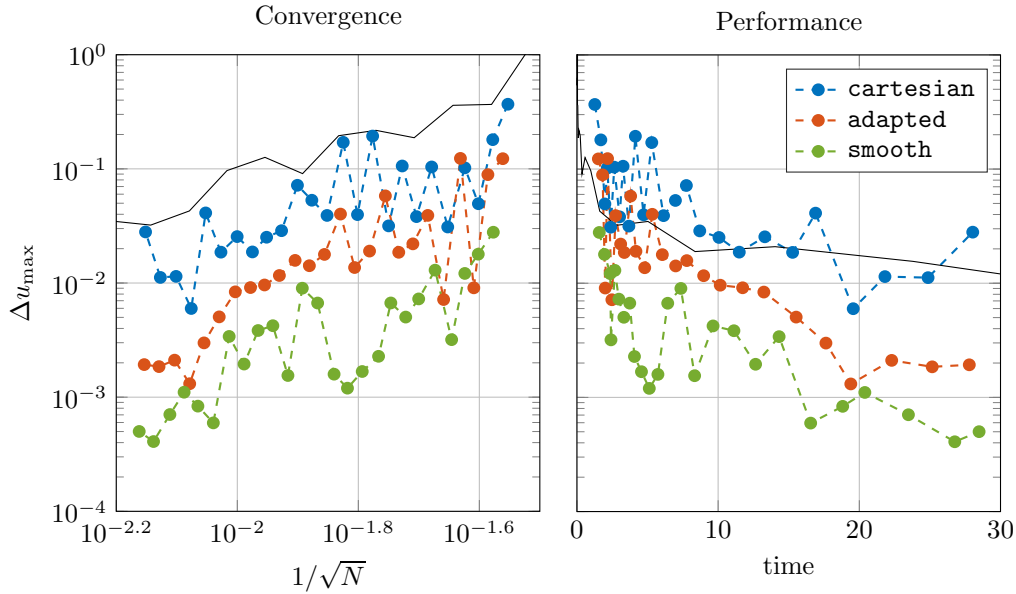
\par
The performance results in Figure \ref{figBSamPut} look very similar to the European case. Numerical experiments show that the same can be stated for the conditioning. 
%
%
%
%
\subsection{Multi-Factor Model}
\par Models with multiple stochastic factors are introduced to capture market features better than the standard Black--Scholes--Merton model. It is well known that the Black--Scholes--Merton framework fails to model heavy tails of return distributions and volatility skews. For instance, the interest for local volatility models started with the work of Dupire \cite{dupire1994pricing} and ever since they have been becoming increasingly popular. In this section, we use one of the most popular stochastic local volatility models known as the Heston model \cite{heston1993closed}.
\par
The dynamics of the Heston models is
\begin{align}
\dif S(t) & =  rS(t)\dif t + \sqrt{V(t)}S(t)\dif W_s(t), \label{qlsvSDE1} \\
\dif V(t) & =  \kappa(\eta-V(t))\dif t + \sigma \sqrt{V(t)}\dif W_v(t), \label{qlsvSDE2}
\end{align}
where $S(t)$ is the underlying asset price, $V(t)$ is its stochastic volatility, $\sigma$ is the constant volatility of volatility, $\kappa$ is the speed of mean reversion of the volatility process, $\eta$ is the mean reversion level, $r$ is the risk-free interest rate,  $W_s(t)$ and $W_v(t)$ are correlated Wiener processes with constant correlation $\rho$, i.e., $\dif W_s(t) \dif W_v(t) = \rho \dif t$.
\par
By applying the It\^{o} lemma and the Feynman--Kac theorem, a PDE for the Heston model reads as
\begin{align}\label{qlsvPDE}
\frac{\partial u}{\partial t}+\mathcal{L}u&=0,\\
u(s,v,T) &= \max(s-K,\ 0),
\end{align}
where
\begin{equation}
\label{eqHSTop}
\mathcal{L}u \equiv \frac{1}{2}vs^2\frac{\p^2 u}{\p s^2} + \rho\sigma v s \frac{\p^2 u}{\p s\p v} + \frac{1}{2}\sigma^2v\frac{\p^2 u}{\p v^2} + rs\frac{\p u}{\p s} + \kappa(\eta-v)\frac{\p u}{\p v} - ru,
\end{equation}
$K$ is the strike price and $s$ and $v$ are deterministic representations of the stochastic asset price and volatility processes, respectively.
\subsubsection{European Call Option under The Heston Model}
\begin{figure}[H]
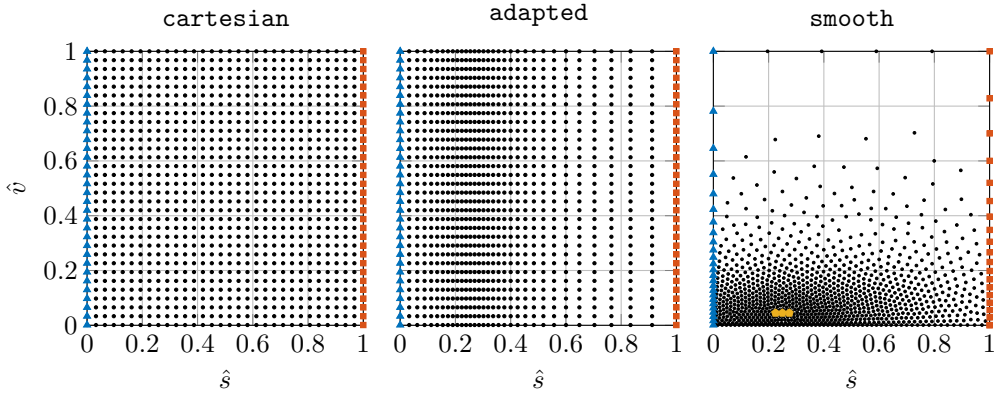

\centering
\input{figs/HSTgrid_cart.tikz}
\hspace{0.25cm}
\input{figs/HSTgrid_adap.tikz}
\hspace{0.25cm}
\input{figs/HSTgrid_smooth.tikz}
\caption{Node layouts on a scaled computational domain used in the numerical experiments. Blue triangles denote the nodes where the close field boundary conditions are applied, red squares denote the nodes where the far field boundary conditions are applied, and yellow pentagons denote the nodes in which the error is measured.}
\label{figHSTgrid}
\end{figure}

We perform an experiment with a European call option where $r=0.03$, $\kappa=2$, $\eta=0.0225$, $\sigma=0.25$, and $\rho=-0.5$, while $K=100$. We choose three evaluation points close to the strike price $K$ at which we compute the option value results
$$\mathbf{x}_{\texttt{HST}} =
\begin{pmatrix}
90 & 0.0225\\
100 & 0.0225\\
110 & 0.0225
\end{pmatrix},$$
The relative location of these points on a scaled domain can be seen in the right most plot of Figure \ref{figHSTgrid} denoted with yellow pentagons. For the \texttt{cartesian} and \texttt{adapted} node layouts, we employ cubic interpolation to approximate the values at the desired points. Then, we use the maximum error from those three points $\Delta u_{\max}$ to measure convergence.
\par
We setup the computational domain such that the far field boundaries are at $s^{\max}=4K$ and $v^{\max}=0.5$. At the points where $s=0$, we set a simple Dirichlet boundary condition $u(\mathbf{x}_{\texttt{CF}},t)=0$, and at the far field $u(\mathbf{x}_{\texttt{FF}},t) = s-K\exp(-rt)$. The points $\mathbf{x}_{\texttt{CF}}$ are marked by blue triangles, while the $\mathbf{x}_{\texttt{FF}}$ points are marked by red squares in the plots of Figure \ref{figHSTgrid}. We leave the volatility boundaries without enforcing any conditions and compute the option values at those points through the RBF-FD approximation, the same way as in the inner domain.
\par
Similarly to the previous examples, we consider three different node layouts for this problem and we name them correspondingly. In this case the \texttt{adapted} node layout is not constructed diagonally, as the discontinuity in the terminal condition of this problem is orthogonal to $s$ axis. This node layout has been adapted to cluster the points close to the strike. In case of the \texttt{smooth} node layout, the nodes are pushed closer to the strike and towards the lower volatility boundary. This is done by choosing the node layout parameters as $P=0.75$, $Q=0.25$, and $G=0$, while $X_1=\hat{K}=\frac{1}{4}$ and $X_2=\frac{1}{2}\ 0.0225$ .  An example of the node layouts where $N\approx1000$ is shown in Figure \ref{figHSTgrid}.
\par
When it comes to performance of the RBF-FD methods under the Heston model, Figure \ref{figHST} shows how important it is to have a smoothly varying density. The figure clearly demonstrates the dominance of the RBF-FD method with the \texttt{smooth} layout, both in accuracy and computational time. In this figure, we can also see how the \texttt{adapted} layout, which was initially designed for basket option problems, starts as a bit better than \texttt{cartesian}, but eventually encounters numerical instabilities as the layout density fails to be smooth enough to support the stencils under the Heston operator. Figure \ref{figHSTcond} shows the \texttt{smooth} node layout as a tremendously better conditioned scheme compared to the other two.
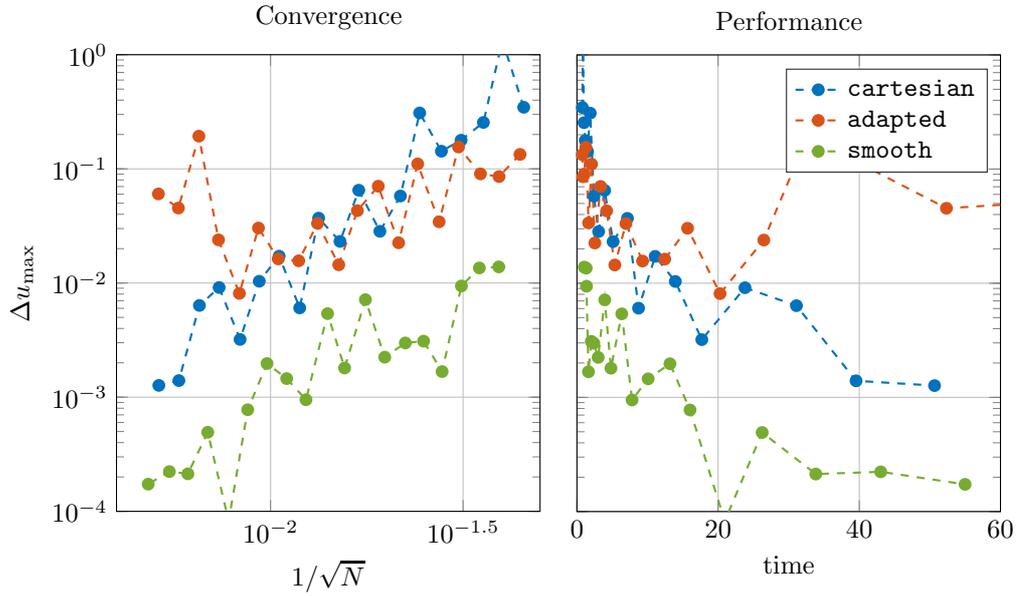
\begin{figure}[H]
\centering
%
%
\definecolor{mycolor1}{rgb}{0.00000,0.44700,0.74100}%
\definecolor{mycolor2}{rgb}{0.85000,0.32500,0.09800}%
\definecolor{mycolor3}{rgb}{0.92900,0.69400,0.12500}%
\definecolor{mycolor4}{rgb}{0.49400,0.18400,0.55600}%
\definecolor{mycolor5}{rgb}{0.46600,0.67400,0.18800}%
\definecolor{mycolor6}{rgb}{0.30100,0.74500,0.93300}%
\begin{tikzpicture}[trim axis left, trim axis right, baseline]

  \begin{axis}[
  grid=major,
  width=0.46\textwidth,
  height=0.5\textwidth,
  at={(0\textwidth,0\textwidth)},
  scale only axis,
  unbounded coords=jump,
  xmode=log,
  xmin=10^-2.4,
  xmax=10^-1.3,
  xlabel={$1/\sqrt{N}$},
  ymode=log,
  ymin=1e-04,
  ymax=1,
  yminorticks=true,
  ytick distance=10^1,
  xminorticks=true,
  ylabel={$\Delta u_{\text{max}}$},
  axis background/.style={fill=white},
  title={Convergence},
  legend pos=north east,
  legend style={legend cell align=left,align=left,draw=white!15!black}
  ]
  \addplot [color=mycolor1,dashed,thick,mark=*,mark options={solid}]
    table[row sep=crcr]{%
    0.0454545454545455	0.346398748156801\\
0.04	1.33044820629064\\
0.0357142857142857	0.255019997546888\\
0.03125	0.178227219495969\\
0.0277777777777778	0.142729018150147\\
0.024390243902439	0.308959906033168\\
0.0217391304347826	0.0578439286099961\\
0.0192307692307692	0.0283782563624921\\
0.0169491525423729	0.0650278539406948\\
0.0151515151515152	0.0231501026675058\\
0.0133333333333333	0.0370277034350455\\
0.0119047619047619	0.0060495170881163\\
0.0105263157894737	0.0171905831628143\\
0.00934579439252336	0.0103615981280099\\
0.00833333333333333	0.00320694517071729\\
0.00735294117647059	0.00913392731230633\\
0.0065359477124183	0.00635791552535059\\
0.00578034682080925	0.00139721158853074\\
0.00512820512820513	0.00126823797955886\\
};
\addlegendentry{\texttt{cartesian}}

  \addplot [color=mycolor2,dashed,thick,mark=*,mark options={solid}]
    table[row sep=crcr]{%
    0.0444554224474387	0.133820278956247\\
0.0392232270276368	0.0856292112892049\\
0.0350931203171798	0.090369004699596\\
0.0307728727448332	0.155384208395635\\
0.0273998312175595	0.0343739195772383\\
0.024098134635594	0.110499212565389\\
0.021506619680967	0.0225053275099956\\
0.0190484829439865	0.07047895387841\\
0.0168073161363204	0.0430606514455767\\
0.0150380190579342	0.014447660925893\\
0.0132453235706504	0.0333019207761485\\
0.0118345267082788	0.0156504676790643\\
0.0104713477072924	0.0162641504102741\\
0.00930242620309913	0.0303133835402676\\
0.00829882662886615	0.00811243236004033\\
0.00732605647520462	0.0238872737801046\\
0.00651469254161754	0.193440125023587\\
0.00576371269475148	0.0453779835972945\\
0.00511510624313881	0.0603334625299334\\
};
\addlegendentry{\texttt{adapted}}
  \addplot [color=mycolor5,dashed,thick,mark=*,mark options={solid}]
    table[row sep=crcr]{%
    0.0391630224993979	0.0138430101053517\\
    0.0348578087187875	0.0135516184012623\\
    0.0313112145542575	0.00941606076014034\\
    0.0278964176325835	0.00167575743301818\\
    0.0249688084719461	0.00309502524629002\\
    0.0223886831419823	0.0029844422465235\\
    0.019799069069658	0.00224080188118769\\
    0.0176363825845535	0.00714867091050664\\
    0.0155718664966075	0.00180234380097222\\
    0.0140677729685715	0.00539301854656138\\
    0.0123607600123336	0.000949533133399072\\
    0.0110123076252695	0.00145444232652281\\
    0.00979075562484947	0.00197096518943329\\
    0.00872373204317023	0.000776197608748586\\
    0.00778428181179587	8.24503420620925e-05\\
    0.00686948649794224	0.000492469356254688\\
    0.00610005630377953	0.000213104786910279\\
    0.00546065035413149	0.00022288213928876\\
    0.00481125224324688	0.000173027587048136\\
    };
    \addlegendentry{\texttt{smooth}}
  \legend{};
\end{axis}
\end{tikzpicture}%
\hspace{0.25cm}
%
%
\definecolor{mycolor1}{rgb}{0.00000,0.44700,0.74100}%
\definecolor{mycolor2}{rgb}{0.85000,0.32500,0.09800}%
\definecolor{mycolor3}{rgb}{0.92900,0.69400,0.12500}%
\definecolor{mycolor4}{rgb}{0.49400,0.18400,0.55600}%
\definecolor{mycolor5}{rgb}{0.46600,0.67400,0.18800}%
\definecolor{mycolor6}{rgb}{0.30100,0.74500,0.93300}%
\begin{tikzpicture}[trim axis left, trim axis right, baseline]

  \begin{axis}[
  grid=major,
  width=0.46\textwidth,
  height=0.5\textwidth,
  at={(0\textwidth,0\textwidth)},
  scale only axis,
  unbounded coords=jump,
  xmin=0,
  xmax=60,
  xlabel={time},
  ymode=log,
  ymin=1e-04,
  ymax=1,
  yticklabels={,,}, 
  yminorticks=true,
  ytick distance=10^1,
  xminorticks=true,
  xmajorgrids,
  ymajorgrids,
  axis background/.style={fill=white},
  title={Performance},
  legend pos=north east,
  legend style={legend cell align=left,align=left,draw=white!15!black}
  ]
  \addplot [color=mycolor1,dashed,thick,mark=*,mark options={solid}]
    table[row sep=crcr]{%
    0.766213179	0.346398748156801\\
  0.843714491	1.33044820629064\\
  1.000230864	0.255019997546888\\
  1.182233613	0.178227219495969\\
  1.496804686	0.142729018150147\\
  1.8819085	0.308959906033168\\
  2.403415455	0.0578439286099961\\
  3.0710868	0.0283782563624921\\
  3.925471032	0.0650278539406948\\
  5.118395308	0.0231501026675058\\
  7.151216989	0.0370277034350455\\
  8.727308534	0.0060495170881163\\
  11.083327937	0.0171905831628143\\
  13.941316404	0.0103615981280099\\
  17.691529733	0.00320694517071729\\
  23.804667244	0.00913392731230633\\
  31.095356083	0.00635791552535059\\
  39.544406166	0.00139721158853074\\
  50.655151689	0.00126823797955886\\
  };
  \addlegendentry{\texttt{cartesian}}

  \addplot [color=mycolor2,dashed,thick,mark=*,mark options={solid}]
    table[row sep=crcr]{%
    0.755159081	0.133820278956247\\
  0.839057786	0.0856292112892049\\
  1.029981858	0.090369004699596\\
  1.26542664	0.155384208395635\\
  1.650761673	0.033739195772383\\
  2.067210959	0.110499212565389\\
  2.540073839	0.0225053275099956\\
  3.345871996	0.07047895387841\\
  4.210087318	0.0430606514455767\\
  5.365412537	0.014447660925893\\
  6.938907576	0.0333019207761485\\
  9.29632043	0.0156504676790643\\
  12.434367058	0.0162641504102741\\
  15.672697822	0.0303133835402676\\
  20.27170759	0.008111243236004033\\
  26.47153582	0.0238872737801046\\
  32.641114337	0.193440125023587\\
  52.343653195	0.0453779835972945\\
  84.749685592	0.0603334625299334\\
  };
  \addlegendentry{\texttt{adapted}}

  \addplot [color=mycolor5,dashed,thick,mark=*,mark options={solid}]
    table[row sep=crcr]{%
    1.0277998273	0.0138430101053517\\
  1.310652145	0.0135516184012623\\
  1.353470544	0.00941606076014034\\
  1.631939132	0.00167575743301818\\
  2.037750709	0.00309502524629002\\
  2.385069449	0.0029844422465235\\
  3.005714275	0.00224080188118769\\
  3.938921052	0.00714867091050664\\
  4.818287775	0.00180234380097222\\
  6.351285278	0.00539301854656138\\
  7.810727765	0.000949533133399072\\
  10.091038613	0.00145444232652281\\
  13.179621808	0.00197096518943329\\
  16.023542842	0.000776197608748586\\
  20.946834195	8.24503420620925e-05\\
  26.239240493	0.000492469356254688\\
  33.837300786	0.000213104786910279\\
  43.037387762	0.00022288213928876\\
  54.97907206	0.000173027587048136\\
  };
  \addlegendentry{\texttt{smooth}}
\end{axis}
\end{tikzpicture}%
\caption{Performance of the RBF-FD method with PHSs for an European call option under the Heston model on different node layouts. The left-hand side plot shows the error against the average node layout density and the right-hand side plot shows the error against the computational time measured in seconds.}
\label{figHST}
\end{figure}
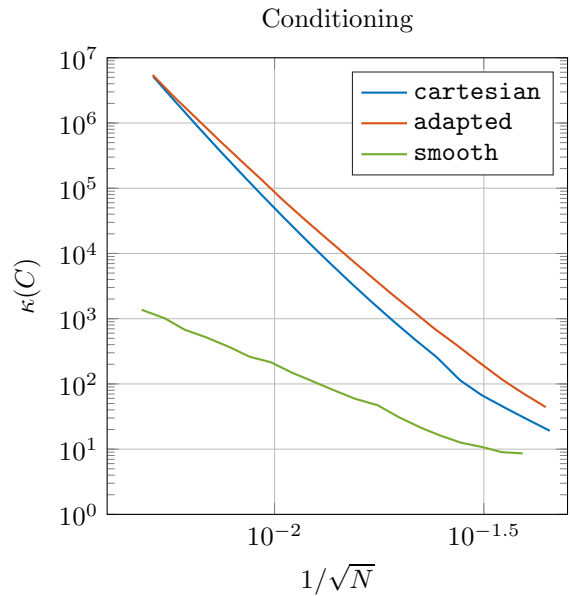
\begin{figure}[H]
\centering
%
%
\definecolor{mycolor1}{rgb}{0.00000,0.44700,0.74100}%
\definecolor{mycolor2}{rgb}{0.85000,0.32500,0.09800}%
\definecolor{mycolor3}{rgb}{0.92900,0.69400,0.12500}%
\definecolor{mycolor4}{rgb}{0.49400,0.18400,0.55600}%
\definecolor{mycolor5}{rgb}{0.46600,0.67400,0.18800}%
\definecolor{mycolor6}{rgb}{0.30100,0.74500,0.93300}%
\begin{tikzpicture}[trim axis left, trim axis right, baseline]

  \begin{axis}[
  grid=major,
  width=0.5\textwidth,
  height=0.5\textwidth,
  at={(0\textwidth,0\textwidth)},
  scale only axis,
  unbounded coords=jump,
  xmode=log,
  xmin=10^-2.4,
  xmax=10^-1.3,
  xlabel={$1/\sqrt{N}$},
  ymode=log,
  ymin=10^0,
  ymax=1e7,
  yminorticks=true,
  ylabel={$\kappa(C)$},
  axis background/.style={fill=white},
  title={Conditioning},
  legend pos=north east,
  legend style={legend cell align=left,align=left,draw=white!15!black}
]
\addplot [color=mycolor1, thick]
  table[row sep=crcr]{%
0.0454545454545455	19.1405320602236\\
0.04	29.0066278865448\\
0.0357142857142857	42.3363571854562\\
0.03125	67.3624302543407\\
0.0277777777777778	113.154920155274\\
0.024390243902439	257.466590231978\\
0.0217391304347826	469.159354401999\\
0.0192307692307692	916.100343578053\\
0.0169491525423729	1877.39838400505\\
0.0151515151515152	3631.19308131796\\
0.0133333333333333	7894.87712879589\\
0.0119047619047619	16044.2669625183\\
0.0105263157894737	35348.7659214015\\
0.00934579439252336	77253.3624772527\\
0.00833333333333333	166913.386369711\\
0.00735294117647059	393325.838893225\\
0.0065359477124183	897649.203726134\\
0.00578034682080925	2157553.99918682\\
0.00512820512820513	5151650.44591735\\
};
\addlegendentry{\texttt{cartesian}}

\addplot [color=mycolor2, thick]
  table[row sep=crcr]{%
0.0444554224474387	44.0683191660782\\
0.0392232270276368	72.0975922759499\\
0.0350931203171798	115.080560576555\\
0.0307728727448332	215.840534186518\\
0.0273998312175595	383.441859654564\\
0.024098134635594	703.680929740761\\
0.021506619680967	1278.00743424322\\
0.0190484829439865	2390.38449739295\\
0.0168073161363204	4718.14224877374\\
0.0150380190579342	8693.18957731747\\
0.0132453235706504	17435.584826762\\
0.0118345267082788	32657.169754367\\
0.0104713477072924	65881.4510789292\\
0.00930242620309913	135797.205887381\\
0.00829882662886615	266823.567349141\\
0.00732605647520462	565181.839789764\\
0.00651469254161754	1169241.62411418\\
0.00576371269475148	2486691.05501432\\
0.00511510624313881	5440266.44649954\\
};
\addlegendentry{\texttt{adapted}}

\addplot [color=mycolor5, thick]
  table[row sep=crcr]{%
0.0391630224993979	8.6019813741126\\
0.0348578087187875	9.00939331682541\\
0.0313112145542575	10.8041566348973\\
0.0278964176325835	12.5484500478369\\
0.0249688084719461	16.0369800725674\\
0.0223886831419823	21.2686849602049\\
0.019799069069658	30.9939349656045\\
0.0176363825845535	47.1226240292348\\
0.0155718664966075	59.2397984163007\\
0.0140677729685715	76.9064177688044\\
0.0123607600123336	109.067909335554\\
0.0110123076252695	148.640869302036\\
0.00979075562484947	215.433965921203\\
0.00872373204317023	260.608807311561\\
0.00778428181179587	369.863743361716\\
0.00686948649794224	516.82401723517\\
0.00610005630377953	678.055225368881\\
0.00546065035413149	1016.44517451785\\
0.00481125224324688	1369.8132962859\\
};
\addlegendentry{\texttt{smooth}}

\end{axis}
\end{tikzpicture}%
\caption{The condition number of the differentiation matrix as a function of the average node density for different node layouts.}
\label{figHSTcond}
\end{figure}
%
%
%
%

\section{Conclusions}
\label{sec4}
\par In this paper we study the benefits of using PHSs and node layouts with smoothly varying density for developing robust and efficient RBF-FD methods for option pricing. We present the improved RBF-FD scheme and successfully apply it to two types of multidimensional PDEs in finance: two-dimensional European call and American put basket options under the Black--Scholes--Merton model, and a European call option under the Heston model. We show numerically that the performance of the method is equally high when it comes to pricing American options compared to the European ones. By studying convergence, computational performance, and conditioning of the discrete systems, we demonstrate the desirable properties of the introduced approaches.
\par
The implemented RBF-FD methods significantly outperformed the standard FD method in the numerical experiments, despite the computational overhead from the differentiation weights. As the computation of the differentiation weights is parallelizable, the performance dominance should be even higher when machines with higher number of cores are used.
\par 
Using PHSs as RBFs, augmented with polynomials, in the RBF-FD approximations, shows to be hassle free as a result of absence of the shape parameter. The PHSs take control of stabilizing the stencils as the degree of the augmented polynomials in the approximation dictates the formal order of the method.
\par 
Although the used smoothly varying density node placing algorithm works only in two-dimensional domains, some recent work has been done to come up with more robust and efficient ways to construct adaptable smooth node layouts in higher dimensions \cite{vlasiuk2017fast}. Research on efficient generation of high-dimensional node layouts is expected to give a significant improvement in performance of the higher-dimensional RBF-FD methods and improve the competitiveness of these methods in different financial applications.
%
%
%
\section*{Acknowledgements}
\par The author is grateful to Natasha Flyer for great discussions on the topic and for sharing the code for generating node layouts. Moreover, gratitude is owed to Lina von Sydow for continuous constructive feedback on the results as well as for proofreading the manuscript.

%
%
%
\bibliographystyle{acm}
\bibliography{paper5}

\begin{thebibliography}{10}

\bibitem{bayona2017role}
{\sc Bayona, V., Flyer, N., Fornberg, B., and Barnett, G.~A.}
\newblock On the role of polynomials in {RBF}--{FD} approximations: {II}.
  {N}umerical solution of elliptic {P}{D}{E}s.
\newblock {\em Journal of Computational Physics 332\/} (2017), 257--273.

\bibitem{bentley1975multidimensional}
{\sc Bentley, J.~L.}
\newblock Multidimensional binary search trees used for associative searching.
\newblock {\em Communications of the ACM 18}, 9 (1975), 509--517.

\bibitem{black73}
{\sc Black, F., and Scholes, M.}
\newblock The pricing of options and corporate liabilities.
\newblock {\em J. Polit. Econ. 81\/} (1973), 637--654.

\bibitem{davydov2011adaptive}
{\sc Davydov, O., and Oanh, D.~T.}
\newblock Adaptive meshless centres and {R}{B}{F} stencils for {P}oisson
  equation.
\newblock {\em Journal of Computational Physics 230}, 2 (2011), 287--304.

\bibitem{dupire1994pricing}
{\sc Dupire, B., et~al.}
\newblock Pricing with a smile.
\newblock {\em Risk 7}, 1 (1994), 18--20.

\bibitem{fasshauer2004using}
{\sc Fasshauer, G.~E., Khaliq, A. Q.~M., and Voss, D.~A.}
\newblock Using meshfree approximation for multi-asset american options.
\newblock {\em Journal of the Chinese Institute of Engineers 27}, 4 (2004),
  563--571.

\bibitem{flyer2016enhancing}
{\sc Flyer, N., Barnett, G.~A., and Wicker, L.~J.}
\newblock Enhancing finite differences with radial basis functions: experiments
  on the {N}avier--{S}tokes equations.
\newblock {\em Journal of Computational Physics 316\/} (2016), 39--62.

\bibitem{flyer2016on}
{\sc Flyer, N., Fornberg, B., Bayona, V., and Barnett, G.~A.}
\newblock On the role of polynomials in {RBF}--{FD} approximations: {I}.
  {I}nterpolation and accuracy.
\newblock {\em Journal of Computational Physics 321\/} (2016), 21--38.

\bibitem{flyer2010rotational}
{\sc Flyer, N., and Lehto, E.}
\newblock Rotational transport on a sphere: Local node refinement with radial
  basis functions.
\newblock {\em Journal of Computational Physics 229}, 6 (2010), 1954--1969.

\bibitem{flyer2012guide}
{\sc Flyer, N., Lehto, E., Blaise, S., Wright, G.~B., and St-Cyr, A.}
\newblock A guide to {R}{B}{F}-generated finite differences for nonlinear
  transport: Shallow water simulations on a sphere.
\newblock {\em Journal of Computational Physics 231}, 11 (2012), 4078--4095.

\bibitem{fornberg2015fast}
{\sc Fornberg, B., and Flyer, N.}
\newblock Fast generation of 2-{D} node distributions for mesh-free {PDE}
  discretizations.
\newblock {\em Computers \& Mathematics with Applications 69}, 7 (2015),
  531--544.

\bibitem{fornberg2011stabilization}
{\sc Fornberg, B., and Lehto, E.}
\newblock Stabilization of {RBF}-generated finite difference methods for
  convective {P}{D}{E}s.
\newblock {\em Journal of Computational Physics 230}, 6 (2011), 2270--2285.

\bibitem{fornberg2013stable}
{\sc Fornberg, B., Lehto, E., and Powell, C.}
\newblock Stable calculation of {G}aussian-based {RBF}--{FD} stencils.
\newblock {\em Comput. Math. Appl. 65}, 4 (Feb. 2013), 627--637.

\bibitem{golbabai2016a}
{\sc Golbabai, A., and Mohebianfar, E.}
\newblock A new stable local radial basis function approach for option pricing.
\newblock {\em Computational Economics\/} (2016), 1--18.

\bibitem{haentjens2015adi}
{\sc Haentjens, T., and in't Hout, K.~J.}
\newblock {ADI} schemes for pricing {A}merican options under the {H}eston
  model.
\newblock {\em Applied Mathematical Finance 22}, 3 (2015), 207--237.

\bibitem{hairer2000solving}
{\sc Hairer, E., N{\o}rsett, S., and Wanner, G.}
\newblock {\em Solving Ordinary Differential Equations {I}. {N}onstiff
  problems}, second~ed.
\newblock Springer-Verlag, Berlin, 2000.

\bibitem{heston1993closed}
{\sc Heston, S.~L.}
\newblock A closed-form solution for options with stochastic volatility with
  applications to bond and currency options.
\newblock {\em The review of financial studies 6}, 2 (1993), 327--343.

\bibitem{hon1999radial}
{\sc Hon, Y.-C., and Mao, X.-Z.}
\newblock A radial basis function method for solving options pricing model.
\newblock {\em Financial Engineering 8}, 1 (1999), 31--49.

\bibitem{ikonen2004operator}
{\sc Ikonen, S., and Toivanen, J.}
\newblock Operator splitting methods for {A}merican option pricing.
\newblock {\em Applied mathematics letters 17}, 7 (2004), 809--814.

\bibitem{ikonen2009operator}
{\sc Ikonen, S., and Toivanen, J.}
\newblock Operator splitting methods for pricing {A}merican options under
  stochastic volatility.
\newblock {\em Numerische Mathematik 113}, 2 (2009), 299--324.

\bibitem{hout2010adi}
{\sc In't~Hout, K., and Foulon, S.}
\newblock {ADI} finite difference schemes for option pricing in the {H}eston
  model with correlation.
\newblock {\em Int. J. Numer. Anal. Model 7}, 2 (2010), 303--320.

\bibitem{kadalbajoo2013application}
{\sc Kadalbajoo, M.~K., Kumar, A., and Tripathi, L.~P.}
\newblock Application of radial basis function with {L}-stable {P}ad{\'e} time
  marching scheme for pricing exotic option.
\newblock {\em Computers \& Mathematics with Applications 66}, 4 (2013),
  500--511.

\bibitem{kadalbajoo2015application}
{\sc Kadalbajoo, M.~K., Kumar, A., and Tripathi, L.~P.}
\newblock Application of the local radial basis function-based finite
  difference method for pricing {A}merican options.
\newblock {\em International Journal of Computer Mathematics 92}, 8 (2015),
  1608--1624.

\bibitem{kadalbajoo2015an}
{\sc Kadalbajoo, M.~K., Kumar, A., and Tripathi, L.~P.}
\newblock An efficient numerical method for pricing option under jump diffusion
  model.
\newblock {\em International Journal of Advances in Engineering Sciences and
  Applied Mathematics 7}, 3 (2015), 114--123.

\bibitem{kansa1990multiquadrics1}
{\sc Kansa, E.~J.}
\newblock Multiquadrics --- {A} scattered data approximation scheme with
  applications to computational fluid-dynamics --- {I} surface approximations
  and partial derivative estimates.
\newblock {\em Computers \& Mathematics with applications 19}, 8-9 (1990),
  127--145.

\bibitem{kansa1990multiquadrics2}
{\sc Kansa, E.~J.}
\newblock Multiquadrics --- {A} scattered data approximation scheme with
  applications to computational fluid-dynamics --- {II} solutions to parabolic,
  hyperbolic and elliptic partial differential equations.
\newblock {\em Computers \& mathematics with applications 19}, 8-9 (1990),
  147--161.

\bibitem{kumar2015numerical}
{\sc Kumar, A., Tripathi, L.~P., and Kadalbajoo, M.~K.}
\newblock A numerical study of {A}sian option with radial basis functions based
  finite differences method.
\newblock {\em Engineering Analysis with Boundary Elements 50\/} (2015), 1--7.

\bibitem{larsson2008multi}
{\sc Larsson, E., {\AA}hlander, K., and Hall, A.}
\newblock Multi-dimensional option pricing using radial basis functions and the
  generalized {F}ourier transform.
\newblock {\em Journal of Computational and Applied Mathematics 222}, 1 (2008),
  175--192.

\bibitem{larsson2013stable}
{\sc Larsson, E., Lehto, E., Heryudono, A., and Fornberg, B.}
\newblock Stable computation of differentiation matrices and scattered node
  stencils based on {G}aussian radial basis functions.
\newblock {\em SIAM Journal on Scientific Computing 35}, 4 (2013),
  A2096--A2119.

\bibitem{merton73}
{\sc Merton, R.~C.}
\newblock Theory of rational option pricing.
\newblock {\em Bell J. Econom. Man. Sci. 4\/} (1973), 141--183.

\bibitem{milovanovic2018radial}
{\sc Milovanovi{\'c}, S., and von Sydow, L.}
\newblock Radial basis function generated finite differences for option pricing
  problems.
\newblock {\em Computers \& Mathematics with Applications 75}, 4 (2018),
  1462--1481.

\bibitem{pettersson2008improved}
{\sc Pettersson, U., Larsson, E., Marcusson, G., and Persson, J.}
\newblock Improved radial basis function methods for multi-dimensional option
  pricing.
\newblock {\em Journal of Computational and Applied Mathematics 222}, 1 (2008),
  82--93.

\bibitem{safdari2015radial}
{\sc Safdari-Vaighani, A., Heryudono, A., and Larsson, E.}
\newblock A radial basis function partition of unity collocation method for
  convection--diffusion equations arising in financial applications.
\newblock {\em Journal of Scientific Computing\/} (2015), 1--27.

\bibitem{salmi2014imex}
{\sc Salmi, S., Toivanen, J., and von Sydow, L.}
\newblock An {IMEX}-scheme for pricing options under stochastic volatility
  models with jumps.
\newblock {\em SIAM Journal on Scientific Computing 36}, 5 (2014), B817--B834.

\bibitem{shcherbakov2016radial}
{\sc Shcherbakov, V.}
\newblock Radial basis function partition of unity operator splitting method
  for pricing multi-asset {A}merican options.
\newblock {\em BIT Numerical Mathematics\/} (2016), 1--23.

\bibitem{shcherbakov2015radial}
{\sc Shcherbakov, V., and Larsson, E.}
\newblock Radial basis function partition of unity methods for pricing vanilla
  basket options.
\newblock {\em Computers \& Mathematics with Applications 71}, 1 (2016),
  185--200.

\bibitem{tolstykh2000using}
{\sc Tolstykh, A.~I.}
\newblock On using {R}{B}{F}-based differencing formulas for unstructured and
  mixed structured--unstructured grid calculations.
\newblock In {\em Proceedings of the 16th IMACS World Congress on Scientific
  Computation, Applied Mathematics and Simulation, Lausanne, Switzerland\/}
  (2000), p.~6.

\bibitem{vlasiuk2017fast}
{\sc Vlasiuk, O., Michaels, T., Flyer, N., and Fornberg, B.}
\newblock Fast high-dimensional node generation with variable density.
\newblock {\em arXiv preprint arXiv:1710.05011\/} (2017).

\bibitem{vonsydow2015benchop}
{\sc von Sydow, L., Josef~H{\"o}{\"o}k, L., Larsson, E., Lindstr{\"o}m, E.,
  Milovanovi{\'c}, S., Persson, J., Shcherbakov, V., Shpolyanskiy, Y.,
  Sir{\'e}n, S., Toivanen, J., et~al.}
\newblock {BENCHOP} --- the {BENCH}marking project in option pricing.
\newblock {\em International Journal of Computer Mathematics 92}, 12 (2015),
  2361--2379.

\bibitem{vonsydow2018benchop}
{\sc von Sydow, L., Milovanovi{\'c}, S., Larsson, E., in~'t Hout, K.,
  Wiktorsson, M., Oosterlee, C.~W., Shcherbakov, V., Wyns, M., Leitao, A.,
  Jain, S., Haentjens, T., and Wald{\'e}n, J.}
\newblock {BENCHOP}--{SLV}: The {BENCH}marking project in option pricing ---
  stochastic and local volatility problems.
\newblock Submited to \emph{International Journal of Computer Mathematics},
  July 2018.

\bibitem{wendland2002fast}
{\sc Wendland, H.}
\newblock Fast evaluation of radial basis functions: Methods based on partition
  of unity.
\newblock In {\em Approximation Theory X: Wavelets, Splines, and
  Applications\/} (2002), Citeseer.

\bibitem{wright2006scattered}
{\sc Wright, G.~B., and Fornberg, B.}
\newblock Scattered node compact finite difference-type formulas generated from
  radial basis functions.
\newblock {\em Journal of Computational Physics 212}, 1 (2006), 99--123.

\end{thebibliography}
%
%
%
%
\end{document}